\documentclass[12pt,preprint,preprintnumbers,bibnotes,showkeys,secnumarabic]{revtex4-1}
\usepackage[T1]{fontenc}
\usepackage[latin9]{inputenc}
\usepackage{color}
\usepackage{amsmath}
\usepackage{amssymb}
\usepackage{graphicx}
\usepackage[unicode=true,
 bookmarks=false,
 breaklinks=false,pdfborder={0 0 1},backref=section,colorlinks=true]
 {hyperref}
\hypersetup{
 linkcolor=blue,anchorcolor=blue,citecolor=blue}

\makeatletter
\usepackage{dcolumn}
\usepackage{bm}
\usepackage{soul}
\usepackage{amsfonts}
\usepackage{ragged2e}

\makeatother

\begin{document}
\title{Single-ion anisotropy effects on the critical behaviors of quantum
entanglement and correlation in the spin-1 Heisenberg chain}
\author{Wanxing Lin$^{1}$$^{,}$$^{2}$}
\thanks{These authors contributed equally to this work.}
\author{Yu-Liang Xu$^{1}$$^{,}$$^{2}$}
\thanks{These authors contributed equally to this work.}
\author{Zhong-Qiang Liu$^{2}$, Chun-Yang Wang$^{1}$$^{,}$$^{2}$}
\author{Xiang-Mu Kong$^{1}$$^{,}$$^{2}$,}
\thanks{Author to whom any correspondence should be addressed. E-mail: kongxm668@163.com~~~~~~~\\
 ORCID iDs~~~~~~~\\
 Wanxing Lin: \textcolor{blue}{0000-0001-9763-6299} ~~~~~~~\\
 Yu-Liang Xu: \textcolor{blue}{0000-0001-9326-7190}~~~~~~~\\
 Zhong-Qiang Liu: \textcolor{blue}{0000-0001-9982-404X} ~~~~~~~\\
 Chun-Yang Wang: \textcolor{blue}{0000-0003-4432-6902} ~~~~~~~\\
 Xiang-Mu Kong: \textcolor{blue}{0000-0002-3891-7629} ~~}
\affiliation{1. School of Physics and Optoelectronic Engineering, Ludong University,
Yantai 264025, China~~~~~~~~~~~~~\\
 2. College of Physics and Physical Engineering, Qufu Normal University,
Qufu 273165, China}
\begin{abstract}
Quantum entanglement and correlations in the spin-1 Heisenberg chain
with single-ion anisotropy are investigated using the quantum renormalization
group method. Negativity and quantum discord (QD) are calculated with
various anisotropy parameters $\bigtriangleup$ and single-ion anisotropy
parameters $D$. We focus on the relations between two abovementioned
physical quantities and on transitions between the Néel, Haldane,
and Large-D phases. It is found that both negativity and QD exhibit
step-like patterns in different phases as the size of the system increases.
Interestingly, the single-ion anisotropy parameter $D$, which can
be modulated using nuclear electric resonance (2020 \textit{Nature}
\textbf{579} 205), plays an important role in tuning the quantum phase
transition (QPT) of the system. Both the first partial derivative
of the negativity and quantum discord with respect to $D$ or $\bigtriangleup$
exhibit nonanalytic behavior at the phase transition points, which
corresponds directly to the divergence of the correlation length.
The quantum correlation critical exponents derived from negativity
and QD are equal, and are the reciprocal of the correlation length
exponent at each critical point. This work extends the application
of quantum entanglement and correlations as tools for depicting QPTs
in spin-1 systems.
\end{abstract}
\maketitle

\section{Introduction}

Entanglement is a peculiar correlation in quantum systems, which is
the fundamental difference between quantum and classical physics \cite{PhysRevLett.70.1895}.
In the past two decades, quantum entanglement has attracted much attention
due to its novel physical properties and its potential applications
in the development of quantum computers and quantum information devices
\cite{PhysRevLett.85.2392}. It has been realized as a crucial resource
in processing and sending quantum information \cite{PhysRevLett.67.661,bouwmeester1997experimental}.
Recently, it has been found that quantum entanglement has a close
relationship with quantum phase transitions (QPTs) and can be widely
exploited for indicating QPTs \cite{PhysRevLett.78.5022,osterloh2002scaling,PhysRevLett.90.227902,PhysRevLett.92.027901}.
Besides quantum entanglement, quantum discord (QD) gives a more common
conception of quantum correlations (QCs), which even occurs in unentangled
systems, and is also a useful measurement tool for depicting QPTs
\cite{PhysRevLett.88.017901,PhysRevA.84.042313}. QPTs occur at absolute
zero temperature, which is induced by the change of an external parameter
or coupling constant. In condensed matter physics, this mechanism
is at the core of relevant quantum phenomena such as superconductivity
and the quantum Hall effect \cite{sachdev2007quantum}. Research into
QPTs is also one of the most interesting topics in strongly correlated
systems, to emerge during the last decade, and investigations of the
relation between QCs and QPTs has attracted much attention recently
\cite{osterloh2002scaling,PhysRevA.66.032110,sachdev2007quantum}.

In the field of strongly correlated systems, various methods are used
extensively to investigate the properties of many-body systems, such
as the renormalization group method \cite{RevModPhys.47.773,Pefeuty1982},
the density matrix renormalization group method \cite{PhysRevLett.69.2863,PhysRevB.53.R10445,PhysRevA.77.012311,PhysRevB.92.195110,PhysRevA.96.032302},
and the tensor renormalization group approach \cite{verstraete2004renormalization,PhysRevLett.106.127202,PhysRevB.100.045110}.
In addition, the quantum renormalization group (QRG) is also a popular
analytic method for investigating the behavior of quantum spin systems.
Quantum entanglement in one- and two-dimensional spin systems has
been investigated using the QRG method, which exhibits nonanalytic
and scaling behaviors in the vicinity of the quantum critical points
\cite{PhysRevB.69.100402,PhysRevA.77.032346,PhysRevB.78.214414,xu2014robust,XU2016217,PhysRevB.98.085136}.
In particular, quantum entanglement in, and QPTs of, spin-1/2 XY models,
including ones with staggered Dzyaloshinskii-Moriya interactions,
were studied using the QRG method. In these cases, the behavior of
the entanglement is closely associated with the quantum critical properties,
and the relation between the entanglement exponent and the correlation
length exponent was obtained in \cite{PhysRevA.83.062309,PhysRevA.84.042302}.
Furthermore, the critical properties of spin systems on a fractal
lattice can also be depicted using entanglement, based on the QRG
method \cite{PhysRevA.95.042327,PhysRevE.97.062134}.

The low-energy behavior of spin-1/2 systems, such as the XY, XYZ,
and XXZ models were extensively investigated in \cite{PhysRevB.69.100402,PhysRevA.77.032346,PhysRevA.83.062309},
whereas QCs in higher spin systems have been less studied until now.
Furthermore, the spin profile of many organic Ni materials with significant
single-ion anisotropy can be described by the spin-1 Heisenberg chain
\cite{PhysRevB.50.9174,PhysRevB.53.15004}, and single-ion anisotropy
interactions can be precisely manipulated using the latest experimental
techniques, such as Nuclear Electric Resonance \cite{Nature2020}.
It is important to investigate the spin-1 Heisenberg chain with a
single-ion anisotropy in the field of condensed matter and quantum
information \cite{PhysRevA.77.012311,PhysRevB.34.6372,PhysRevB.67.104401,PhysRevB.84.220402,Phys.Rev.A.97.042318}.
In previous work, the QRG-flow equations and phase diagram of the
spin-1 Heisenberg chain were obtained using the QRG method \cite{Langari_2013}.
The dynamical spin excitations of this model were also investigated
using quantum Monte Carlo simulations and stochastic analytic continuation
\cite{PhysRevB.103.024403}. To the best of the authors' knowledge, the effects of single-ion anisotropy on the critical behavior
of quantum entanglement and correlations in the spin-1 Heisenberg
chain have been investigated systematically except in the present
work.

In this work, the QCs and QPTs of the spin-1 Heisenberg chain with
single-ion anisotropy are investigated using the QRG method. Both
the calculated negativity and QD are affected by the easy-axis anisotropy
and the single-ion anisotropy parameters. The single-ion anisotropy
can effect the negativity and QD by favoring the alignment of spins.
For the given values of the anisotropy or single-ion anisotropy parameters,
both negativity and QD exhibit step-like patterns in different phases,
which are separated by the phase transition points as the size of
the system increases. Furthermore, the first partial derivative of
the negativity and the QD with respect to the anisotropy or single-ion
anisotropy parameters show nonanalytic behavior with a scaling relation
at the phase transition points. Besides, it is found that negativity
and QD depict the QPT in slightly different ways. This paper is organized
as follows. In Sec. II, the spin model and the QRG method are introduced.
In Sec. III, the entanglement and QD between two blocks are investigated.
We discuss the nonanalytic and the scaling behaviors of the entanglement
and QD in Sec. IV, and summarize in Sec. V.

\section{Model and quantum renormalization group method}

The Hamiltonian of the spin-1 Heisenberg chain with a single-ion anisotropy
is given by
\begin{equation}
H=J\sum_{i=1}^{L}[S_{i}^{x}S_{i+1}^{x}+S_{i}^{y}S_{i+1}^{y}+\bigtriangleup S_{i}^{z}S_{i+1}^{z}+D(S_{i}^{z})^{2}],\tag{1}\label{1}
\end{equation}
where $S^{\alpha}(\alpha=x,y,z)$ are spin-1 operators, $J>0$ is
the antiferromagnetic nearest-neighbor interaction, and $\bigtriangleup$
characterizes the easy-axis anisotropy. The single-ion anisotropy
parameter $D$ can be adjusted in an experiment by utilising the latest
developments in nuclear electric resonance \cite{Nature2020}. The
phase diagram of Hamiltonian Eq.(\ref{1}) is well established \cite{PhysRevB.34.6372,PhysRevB.67.104401,PhysRevB.84.220402}.
Here, we focus our attention on phase transitions among the Néel,
Haldane, and large-D phases for $\bigtriangleup\geq0$. Most parts
of the phase diagram are determined accurately. However, it is difficult
to accurately determine the tri-critical point using numerical analysis.

\begin{figure}[ht]
\centering\includegraphics[width=135mm]{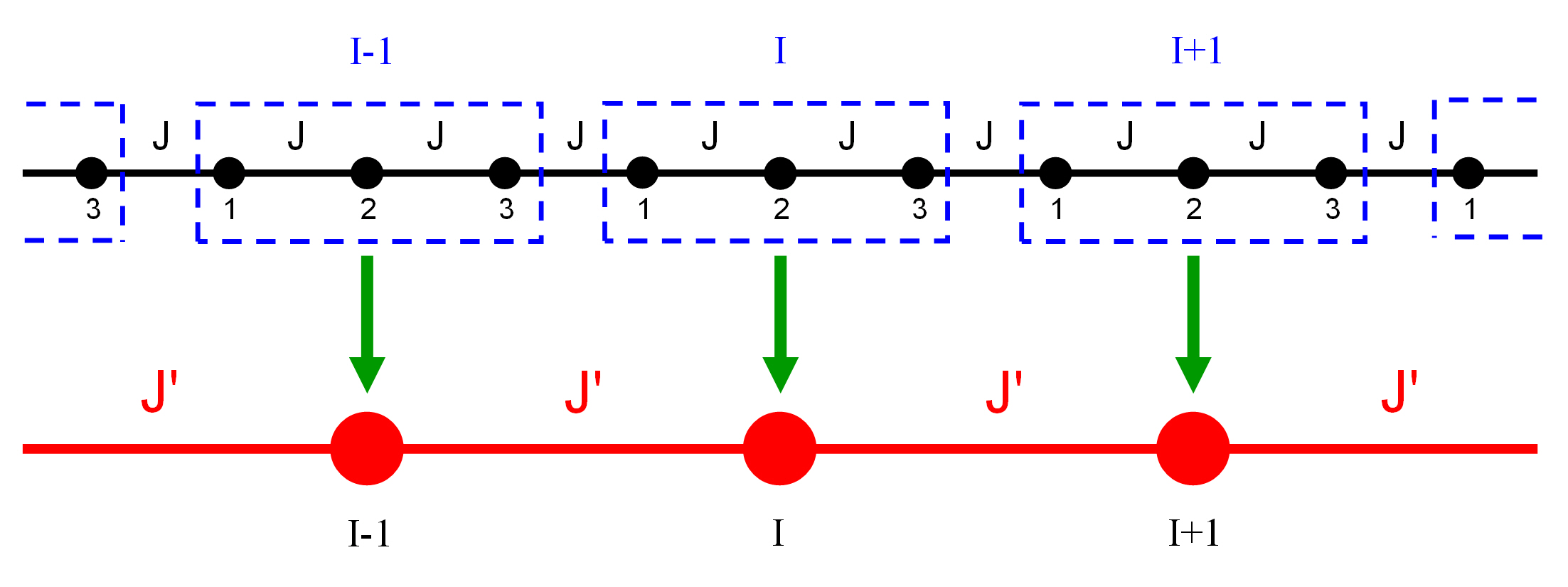}

\caption{Sketch of the renormalization process for the spin-1 chain. The black
(red) dots represent the initial (effective) spins, and the dashed
blue rectangles represent blocks of three spins \cite{Langari_2013}.}
\label{Figure 1}
\end{figure}

\begin{figure}[ht]
\centering\includegraphics[width=86mm]{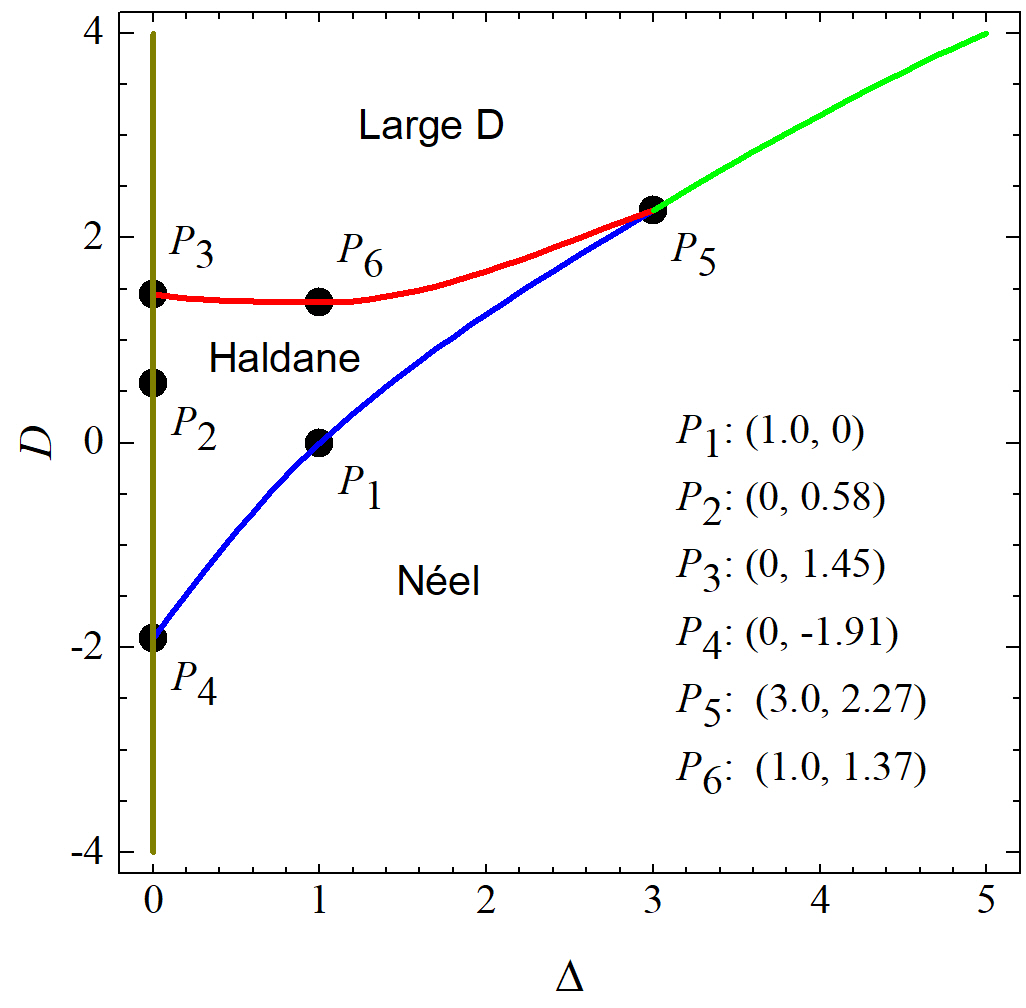}

\caption{Sketch of the phase diagram for the spin-1 Heisenberg model with single-ion
anisotropy. The color lines denote the boundaries between different
phases, and the black dots represent the phase transition points $P_{1}$-$P_{6}$
\cite{Langari_2013}.}
\label{Figure 2}
\end{figure}

The quantum fidelity and QPT of the model were investigated using
the QRG method, based on Kadanoff's block approach \cite{Langari_2013}.
The general idea of the QRG method is to keep the most important degrees
of freedom and integrate out the rest by an iterative procedure. In
this work, the three sites (marked as 1-2-3) of the spin chain are
considered as a block, as shown in figure \ref{Figure 1}, which maps
the initial Hamiltonian into a renormalized Hamiltonian defined by
the set of renormalized couplings $\left(J^{^{\prime}},\bigtriangleup^{^{\prime}},D^{^{\prime}}\right)$.
The relations between the original and renormalized coupling constants
are
\begin{align}
J^{^{\prime}}=\left(X_{ren}\right)^{2}J,\tag{2}\label{2}
\end{align}
\begin{align}
\bigtriangleup^{^{\prime}}=\left(\frac{Z_{ren}}{X_{ren}}\right)^{2}\bigtriangleup,\tag{3}\label{3}
\end{align}
\begin{align}
D^{^{\prime}}=\frac{E_{1}-E_{0}}{\left(X_{ren}\right)^{2}},\tag{4}\label{4}
\end{align}
where $X_{ren}$ and $Z_{ren}$ are the renormalization coefficients,
$E_{0}$ is the ground-state energy of the block Hamiltonian and $E_{1}$
is the first excited-state energy. The first excited-state energy
is doubly degenerate. The explicit form of the renormalized couplings
and the details of the renormalization procedure are presented in
the appendix A. The renormalization of the couplings generates the
flow of the couplings, which in turn determines the quantum phase
diagram and the ground-state properties of the model. The analysis
of the QRG-flow Eqs.(\ref{2})(\ref{3})(\ref{4}) gives a clear picture
of the topography in the ground-state phase diagram. A sketch of the
phase diagram is given in figure \ref{Figure 2}. The QRG-flow includes
two types of fixed points $(\bigtriangleup,D)$, $P_{1}:(1.0,0)$,
$P_{2}:(0,0.58)$, $P_{3}:(0,1.45)$, $P_{4}:(0,-1.91)$, and $P_{5}:(3.0,2.27)$
are fixed points and two others are for extremely large couplings,
namely $\left(\infty,0\right)$ and $\left(-\infty,\infty\right)$.
In particular, both $P_{3}:(0,1.45)$ and $P_{4}:(0,-1.91)$ are quantum
critical points, while $P_{5}:(3.0,2.27)$ is the tri-critical point
\cite{Langari_2013}.

\section{Negativity and QD analysis}

There are many measures for the pairwise entanglement and QC \cite{PhysRevLett.78.5022,PhysRevLett.80.2245,PhysRevA.65.032314,PhysRevA.84.042313}.
Here, we investigate the ground-state entanglement and QC between
two blocks of the spin-1 Heisenberg chain using the measures of negativity
and QD, and demonstrate how they vary as the size of the blocks increase.
Consider the ground state $\left\vert \phi_{0}\right\rangle $ of
a block and define the pure-state density matrix
\begin{equation}
\rho=\left\vert \phi_{0}\right\rangle \left\langle \phi_{0}\right\vert .\tag{5}\label{5}
\end{equation}
Because negativity measures the pairwise entanglement, the degrees
of freedom for one site in the block should be traced out. Generally,
the degrees of freedom of site 2 are traced so that the reduced density
matrix for sites 1 and 3, i.e., $\rho_{13}$, can be obtained. The
negativity of the partial transpose gives a sufficient condition for
the entanglement of spin-1 particles. The negativity of sites 1 and
3 is defined as \cite{PhysRevA.65.032314}
\begin{equation}
N_{13}=\sum_{i}\left\vert \mu_{i}\right\vert ,\tag{6}\label{6}
\end{equation}
where $\mu_{i}$ is the negative eigenvalue of $\rho_{13}^{T_{3}}$,
and $T_{3}$ denotes the partial transpose with respect to the third
spin. The value of $N_{13}$ varies in the range from $0$ to $1$.
If $N_{13}=0$ or $1$, the system is unentangled or entangled maximally,
respectively. Other values correspond to a partially entangled state
\cite{PhysRevA.70.032326}.

QD can be applied to quantify QCs of the system, and is defined by
the formula for mutual information. The quantum mutual information
(QMI) of a bipartite quantum state $\rho_{AB}$ is \cite{PhysRevA.72.032317}
\begin{equation}
I(\rho_{AB})=S(\rho_{A})+S(\rho_{B})-S(\rho_{AB}),\tag{7}\label{7}
\end{equation}
where $S\left(\rho\right)=-Tr\rho Log\rho$ is the von Neumann entropy
of the state $\rho$. The classical correlation is defined in an alternative
version of the mutual information as \cite{PhysRevB.93.184428}
\begin{equation}
J(\rho_{AB})=S(\rho_{A})-min_{\left\{ E_{k}^{B}\right\} }\sum_{k}p_{k}S(\rho_{A}|k),\tag{8}\label{8}
\end{equation}
where the minimum is taken over all possible positive operator-valued
measures (POVMs) $\left\{ E_{k}^{B}\right\} $ on subsystem B with
$p_{k}=Tr\left(E_{k}^{B}\rho_{AB}\right)$ and $\rho_{A}|k=Tr_{B}\left(E_{k}^{B}\rho_{AB}\right)/p_{k}$.
The functions $I(\rho_{AB})$ and $J(\rho_{AB})$ quantify the total
correlation and classical correlation, respectively, and the QD measures
the difference between the two \cite{PhysRevLett.88.017901}:
\begin{equation}
QD(\rho_{AB})=I(\rho_{AB})-J(\rho_{AB}).\tag{9}\label{9}
\end{equation}
QD is considered an effective measure of the QCs of a system, and
we elaborate this numerically in the present work, using the random
unitary matrix method. This allows us to find the minimum over all
POVMs efficiently \cite{PhysRevA.84.042313,PhysRevB.93.184428,PhysRevLett.88.017901}.
As in the negativity defined above, the subscripts $A$ and $B$ in
Eqs.(\ref{7})(\ref{8})(\ref{9}) indicate the sites 1 and 3 in the
spin block, respectively.

\begin{figure}
\centering\includegraphics[width=82mm]{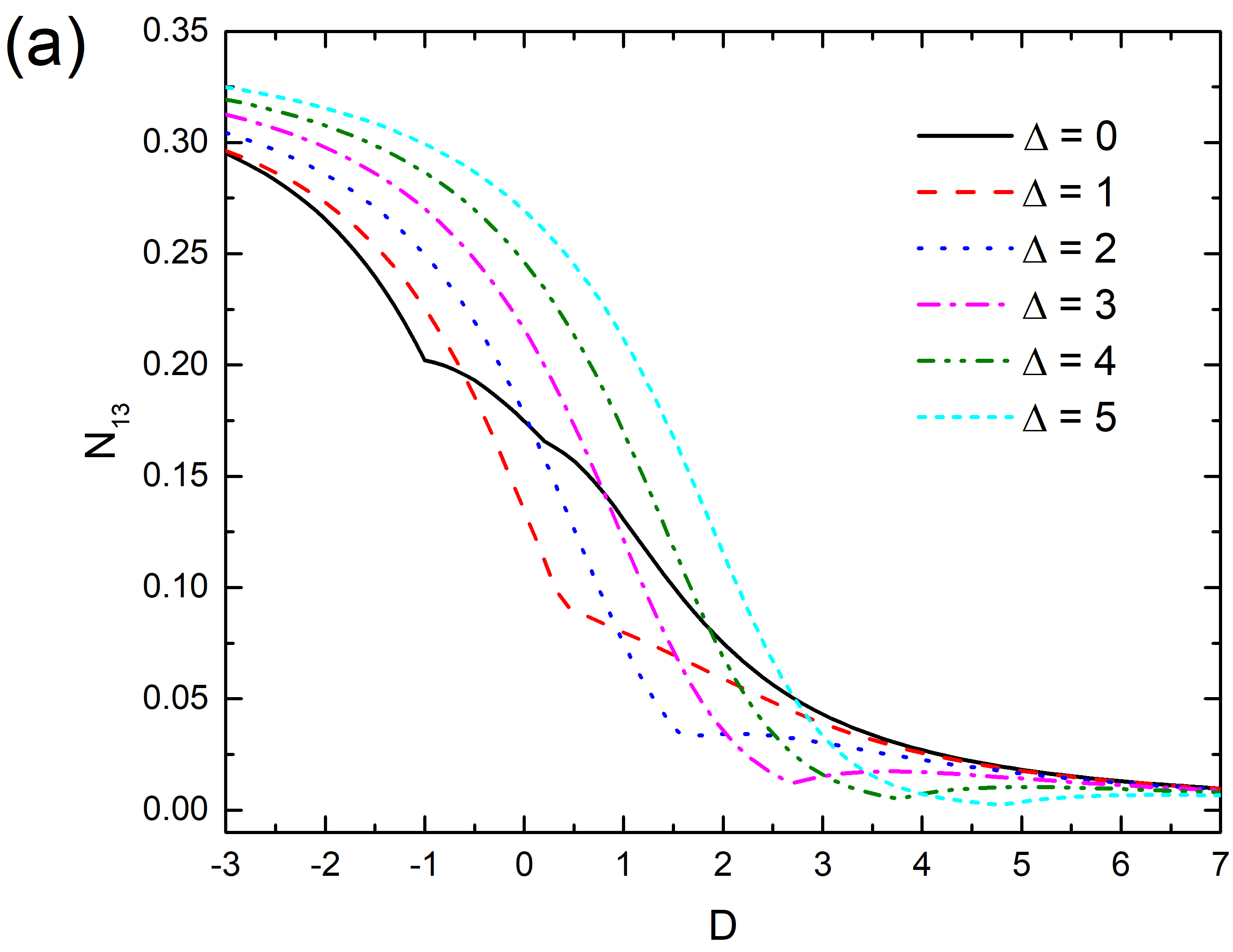}\includegraphics[width=81mm]{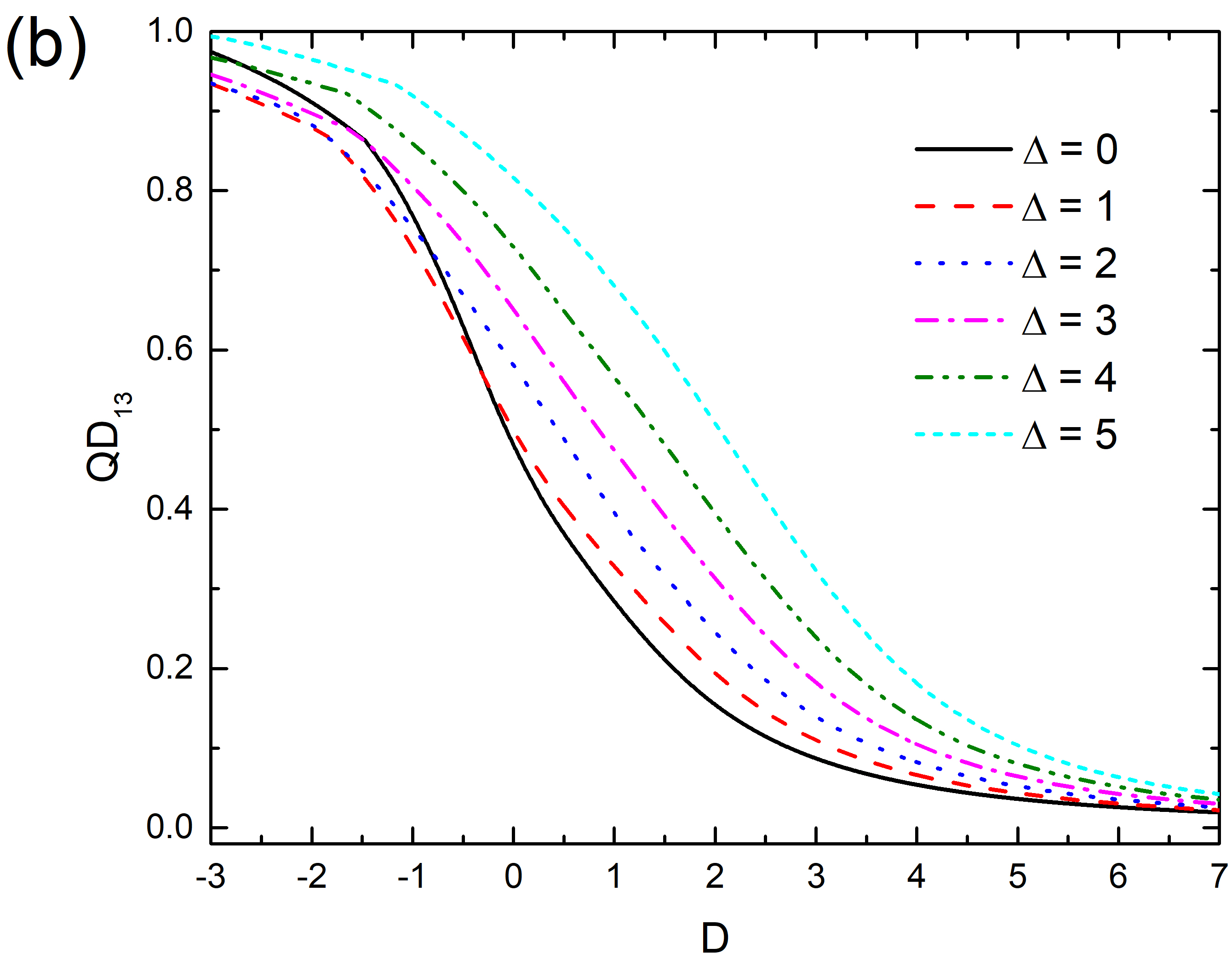}

\caption{(a) Negativity and (b) QD between the first and third sites of the
three-site model in terms of the single-ion anisotropy $D$ for different
values of easy-axis anisotropy $\Delta$.}
\label{Figure 3}
\end{figure}

The numerical calculations indicate that both the negativity and the
QD are influenced by the anisotropy parameter $\bigtriangleup$ and
the single-ion anisotropy parameter $D$. For the three-site model,
we plot the negativity $N_{13}$ versus $D$ for different values
of $\bigtriangleup$, as shown in figure \ref{Figure 3}(a). The negativity
is a decreasing function of the single-ion anisotropy $D$, regardless
of the value of $\bigtriangleup$. In other words, the single-ion
anisotropy suppresses the entanglement by favoring of the alignment
of spins. As the single-ion anisotropy $D$ increases, the probability
of the spin in the block tending to the direction of $\left\vert 0\right\rangle $
is increased. In the limit $D\rightarrow\infty$, the system turns
into a separable state, $\left\vert 000\right\rangle $. Furthermore,
when $D$ is small, $\bigtriangleup$ enhances the entanglement of
the system when $\bigtriangleup\gg0$, while it suppresses the entanglement
as $D$ increases. The QD of the three-site model $QD_{13}$ has a
similar tendency, as shown in figure \ref{Figure 3}(b). The quantum
mutual information of the three-site model $QMI_{13}$ also has a
similar tendency, as shown in figure S1(a) of appendix B. The curves
of both $QD_{13}$ and $QMI_{13}$ are smoother than the curve of
$N_{13}$.

\begin{figure}
\centering\includegraphics[width=82mm]{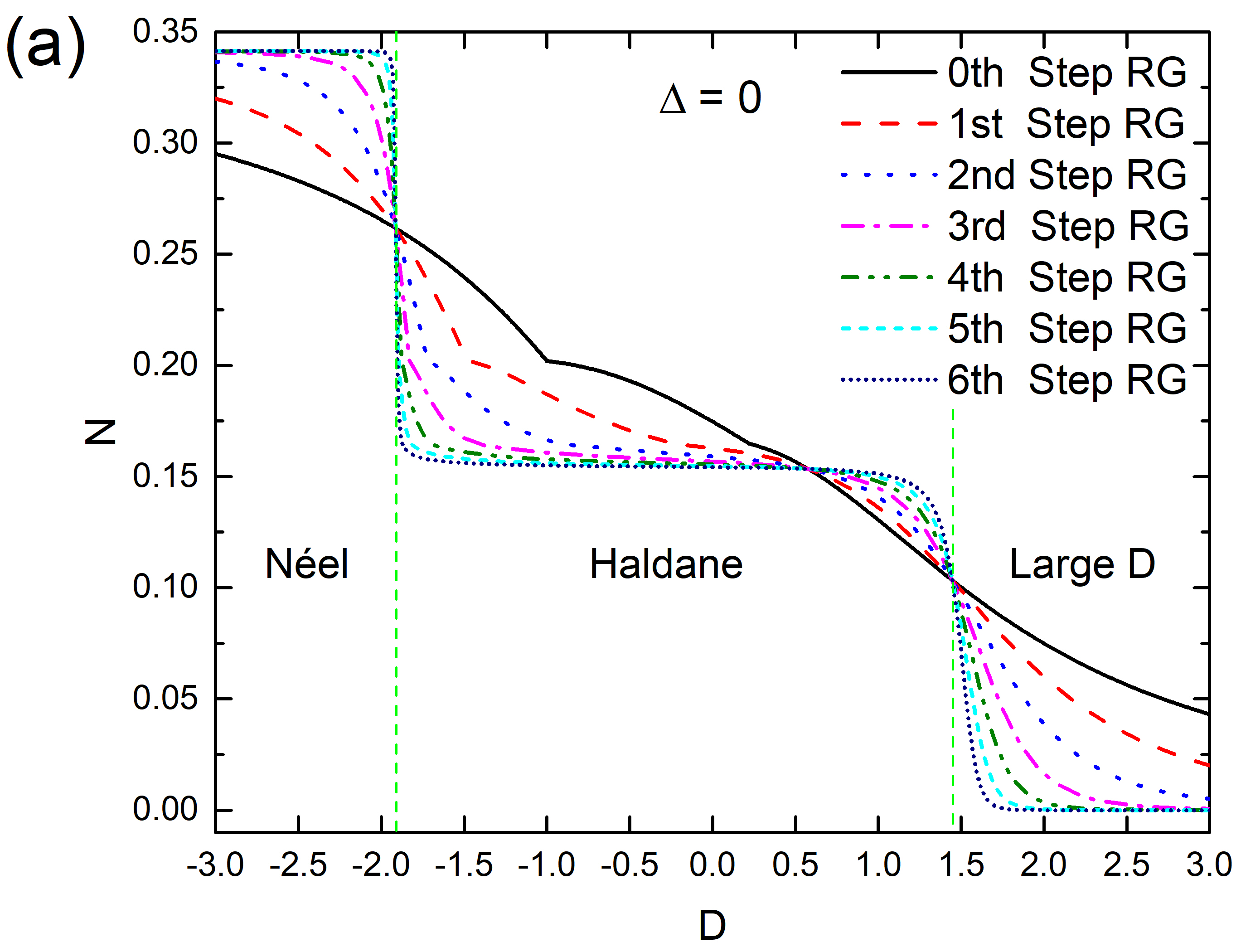}\includegraphics[width=81mm]{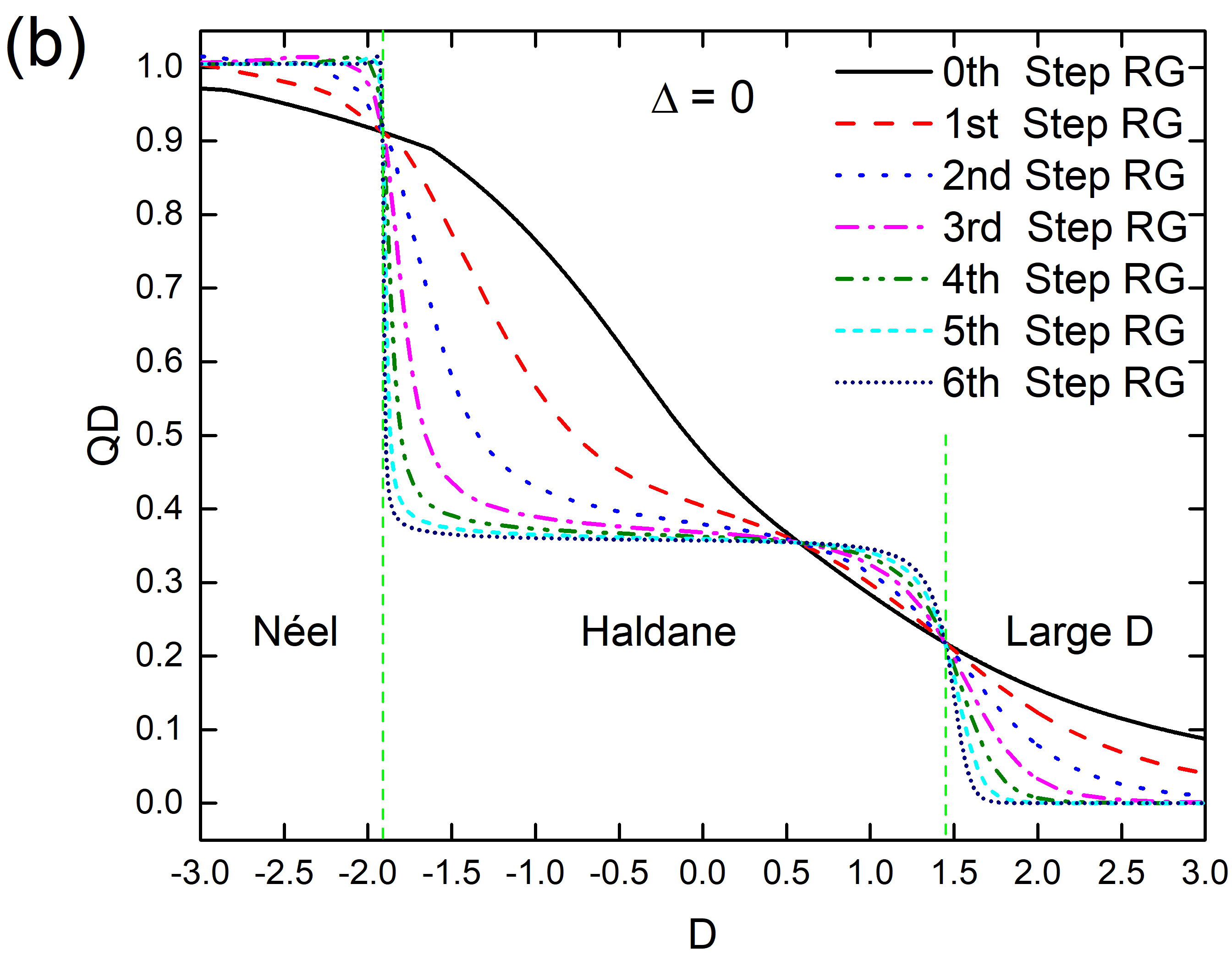}

\caption{(a) Negativity and (b) QD in terms of QRG iterations at $\Delta=0$.
Each phase is labelled by the black text, and separated by the dashed
green lines at $D=-1.91$ and $D=1.45$, respectively.}
\label{Figure 4}
\end{figure}

\begin{figure}
\centering\includegraphics[width=82mm]{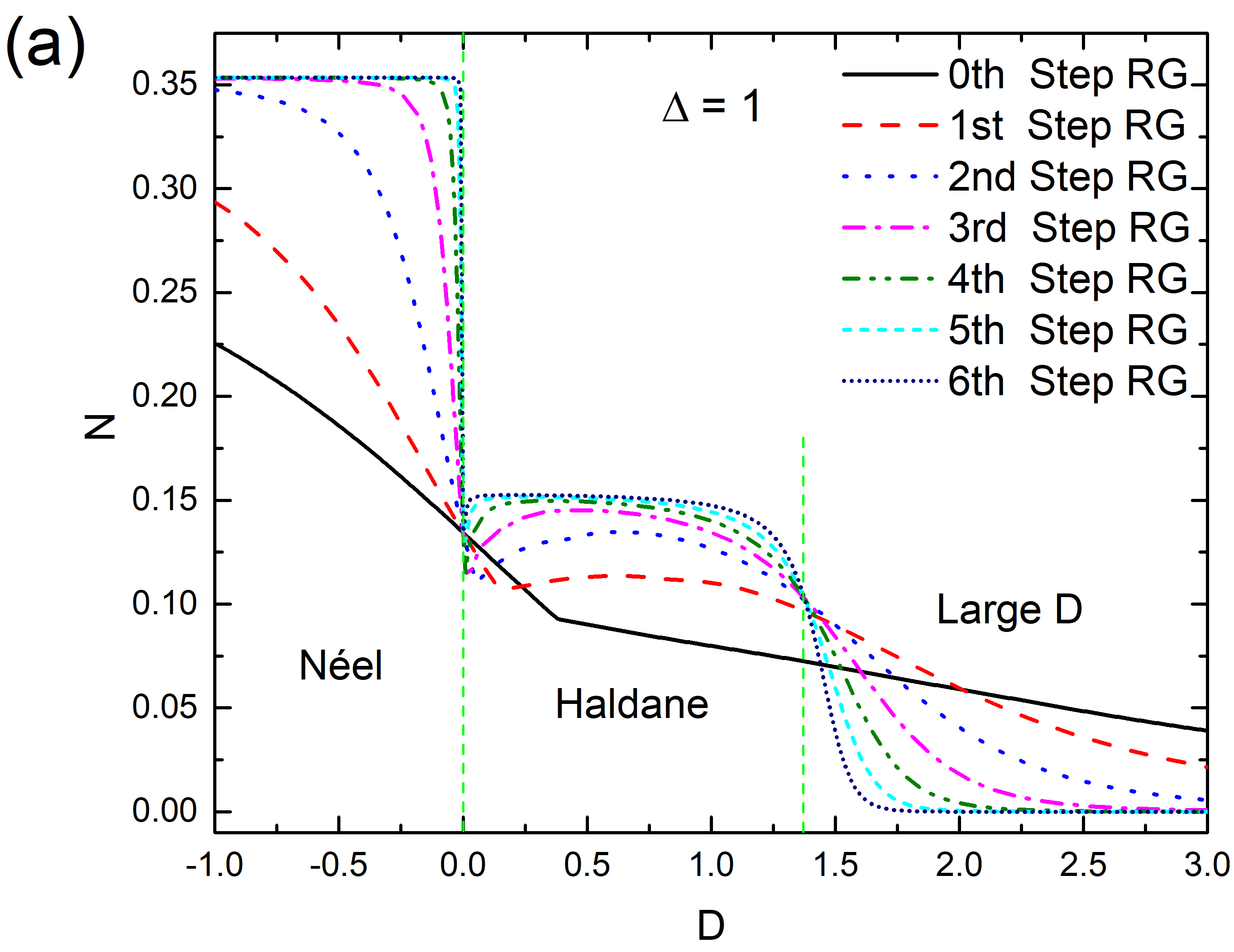}\includegraphics[width=81mm]{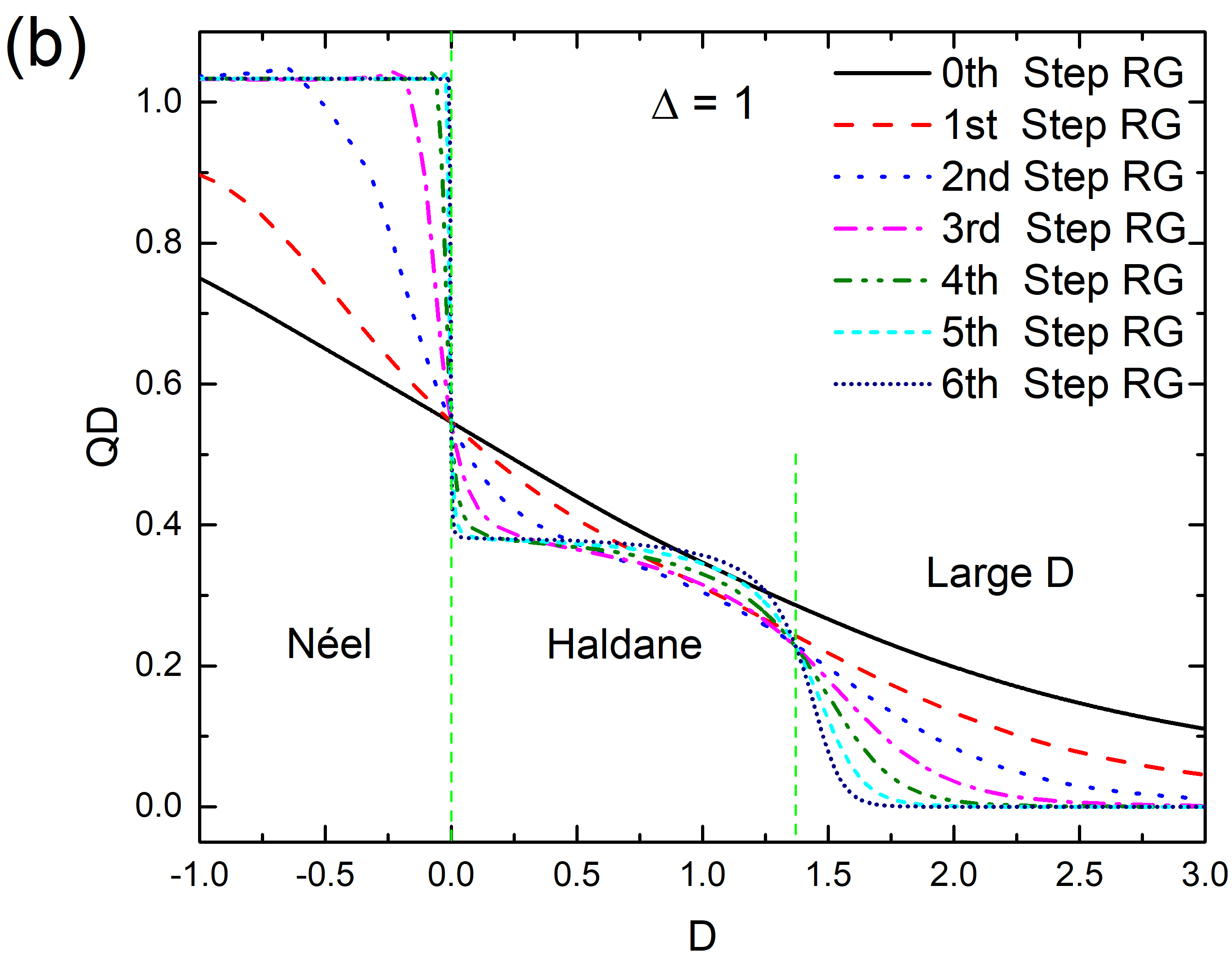}

\caption{(a) Negativity and (b) QD in terms of QRG iterations at $\Delta=1$.
Each phase is labelled by the black text, and separated by the dashed
green lines at $D=0$ and $D=1.37$, respectively.}
\label{Figure 5}
\end{figure}

\begin{figure}
\centering\includegraphics[width=82mm]{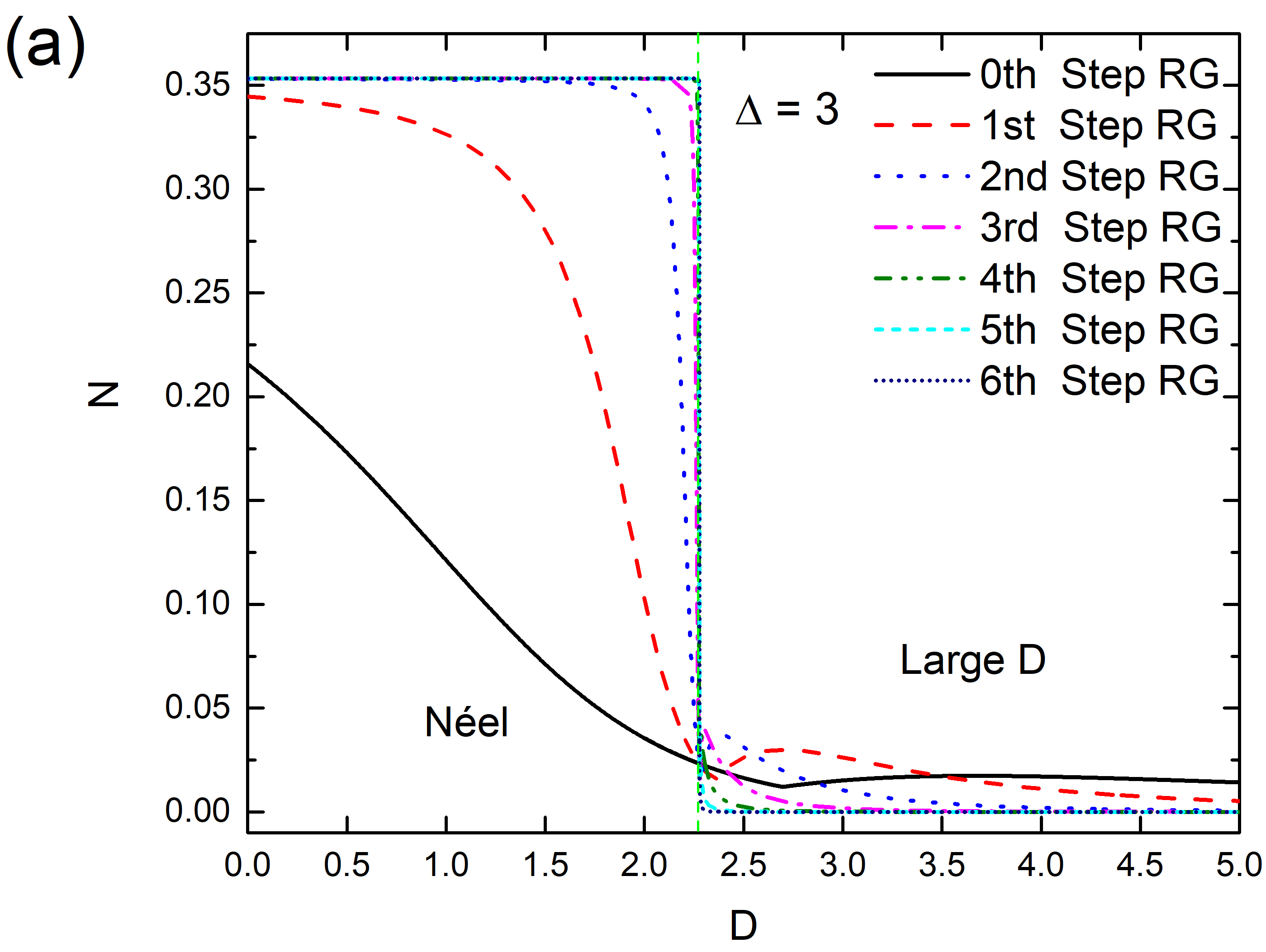}\includegraphics[width=79mm]{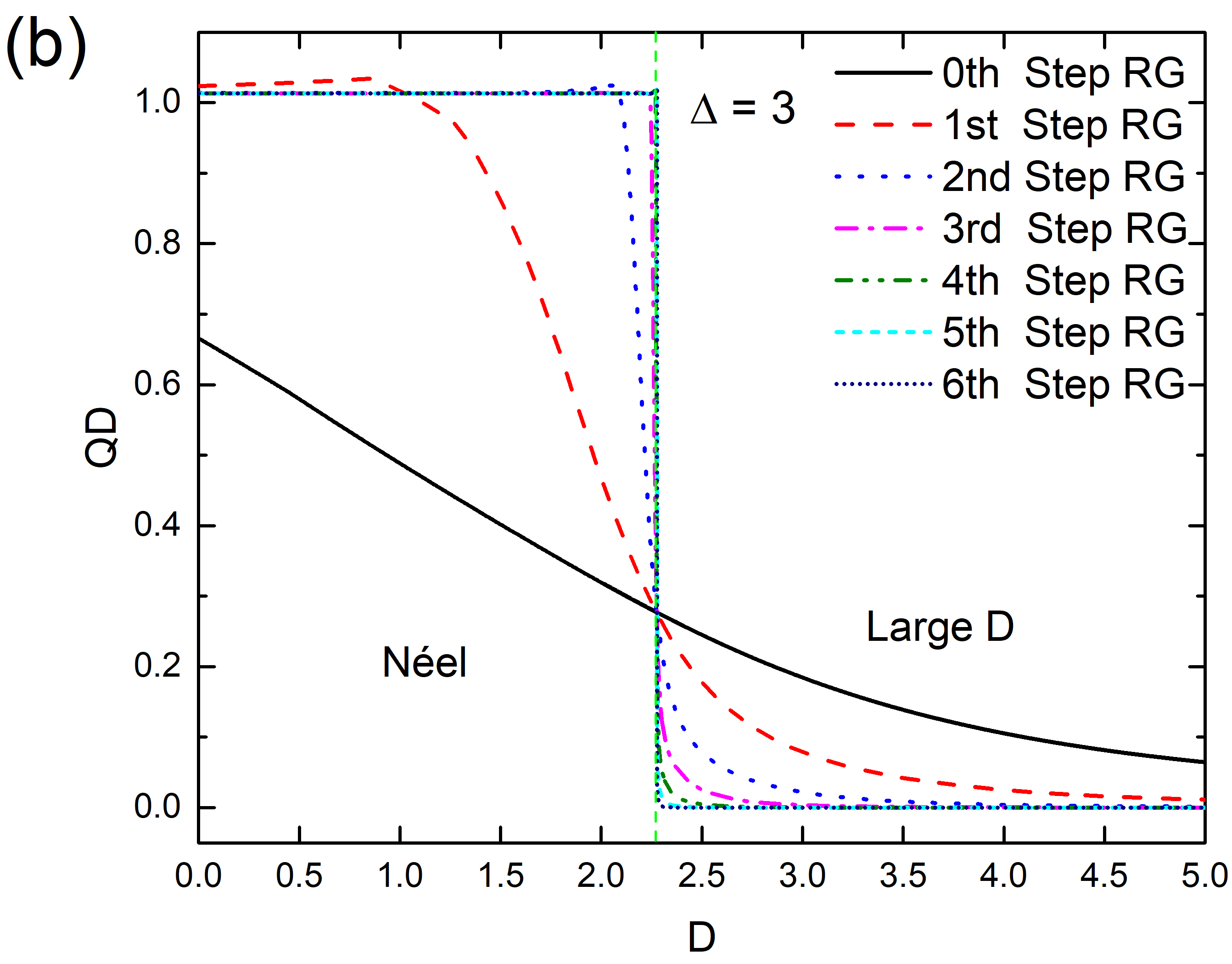}

\caption{(a) Negativity and (b) QD in terms of QRG iterations at $\Delta=3$.
Each phase is labelled by the black text, and separated by the dashed
green line at $D=2.27$.}
\label{Figure 6}
\end{figure}

By combining the negativity with the QRG relations of the renormalized
coupling constants, the entanglement for the large-sized system is
calculated. A plot of the negativity ($N$) versus $D$, with fixed
$\bigtriangleup=0$, is given in figure \ref{Figure 4}(a). The curves
of negativity versus different steps of QRG cross each other at the
fixed points. As the scale of the system increases, the negativity
exhibits three step-like patterns in different phases separated by
the phase transition points, as shown in the dashed green lines of
the figure. The system driven by the single-ion anisotropy $D$ undergoes
a transition from the Néel phase to the Haldane phase with an Ising
transition at the critical point $P_{4}:(0,-1.91)$ \cite{PhysRevA.81.032304}.
This turns the largest negativity into a smaller saturated value,
which then vanishes in the large-D phase through a Gaussian transition
at $P_{3}:(0,1.45)$. Previous works have shown that the Gaussian
transition between the Haldane phase and large-D phase is a symmetry-protected
topological phase transition. There is a lack of a local order parameter,
and the critical exponents of the Gaussian transitions change continuously
along the critical lines \cite{PhysRevB.34.6372,PhysRevA.77.012311,PhysRevB.84.220402}.
At these fixed points, the system exhibits QCs because the negativity
is a nonzero constant. The evolution of QD versus $D$ with $\bigtriangleup=0$
exhibits similar behavior, as shown in figure \ref{Figure 4}(b),
as does the evolution of quantum mutual information ($QMI$) versus
$D$ with $\bigtriangleup=0$ as shown in figure S1(b) of appendix
B. This shows that quantum mutual information can also be used to
depict the QPTs. All curves of negativity, QD, and quantum mutual
information plotted against different steps of the QRG cross each
other at the fixed points $P_{2}:(0,0.58)$. Amazingly, the QD curves
present light humps at the critical point $P_{4}:(0,-1.91)$ as the
size of the system increases, which differs from the behavior of negativity
and quantum mutual information. For a fixed value of $\bigtriangleup=1$,
the curves of negativity versus different steps of the QRG also cross
each other at the fixed points and present step-like patterns in different
phases separated by the fixed points as the system increases, as shown
in figure \ref{Figure 5}(a). The system undergoes an Ising transition
from the Néel phase to the Haldane phase at the fixed point $P_{1}:(1.0,0)$,
which leads to the negativity becoming much smaller. Due to the alignment
of spins, the negativity vanishes altogether in the large-D phase
through a Gaussian transition at $P_{6}:(1.0,1.37)$. These negativities
change smoothly as the single-ion anisotropy $D$ varies at the Gaussian
transition points within SIX steps of RG iteration. The evolution
of QD for $\bigtriangleup=1$ exhibits similar behavior, as shown
in figure \ref{Figure 5}(b). Interestingly, both negativity and QD
only develop two step-like patterns for $\bigtriangleup=3$, as shown
in figures \ref{Figure 6}(a) and (b). With increasing $D$, the system
undergoes QPTs from the Néel phase to the Haldane phase and to the
large-D phase at $\ensuremath{P_{5}:(3.0,2.27)}$ \cite{PhysRevA.81.032304}.
The values of negativity and QD can express the entanglement and correlation
strength of the tri-critical point $P_{5}:(3.0,2.27)$, respectively.
It is obvious that the system exhibits the same entanglement (or QC)
at the tri-critical point as the size of the system increases, which
is caused by the divergence of correlation length.

To compare the two, the negativity and QD are also analyzed by tuning
$\bigtriangleup$ but fixing $D=0$. As the scale of the system increases,
both the negativity and QD develop two step-like patterns separated
by $P_{1}:(1.0,0)$, as shown in figure S2(a) and (b) of appendix
B, respectively. In the thermodynamic limit, both negativity and QD
jump to larger stable values as the system transforms from the Haldane
phase to the Néel phase. The three-site model can describe the infinite
spin-1 chain with renormalized coupling constants. At these critical
points, quantum fluctuations play an important role and destroy any
long-range order of the system. The negativity and QD show a clear
drop at the Néel--Haldane and Néel--large-D phase transition points
as the system reaches $3^{3}$ sites.

\section{Nonanalytic and scaling behavior}

\begin{figure}
\centering\includegraphics[width=82mm]{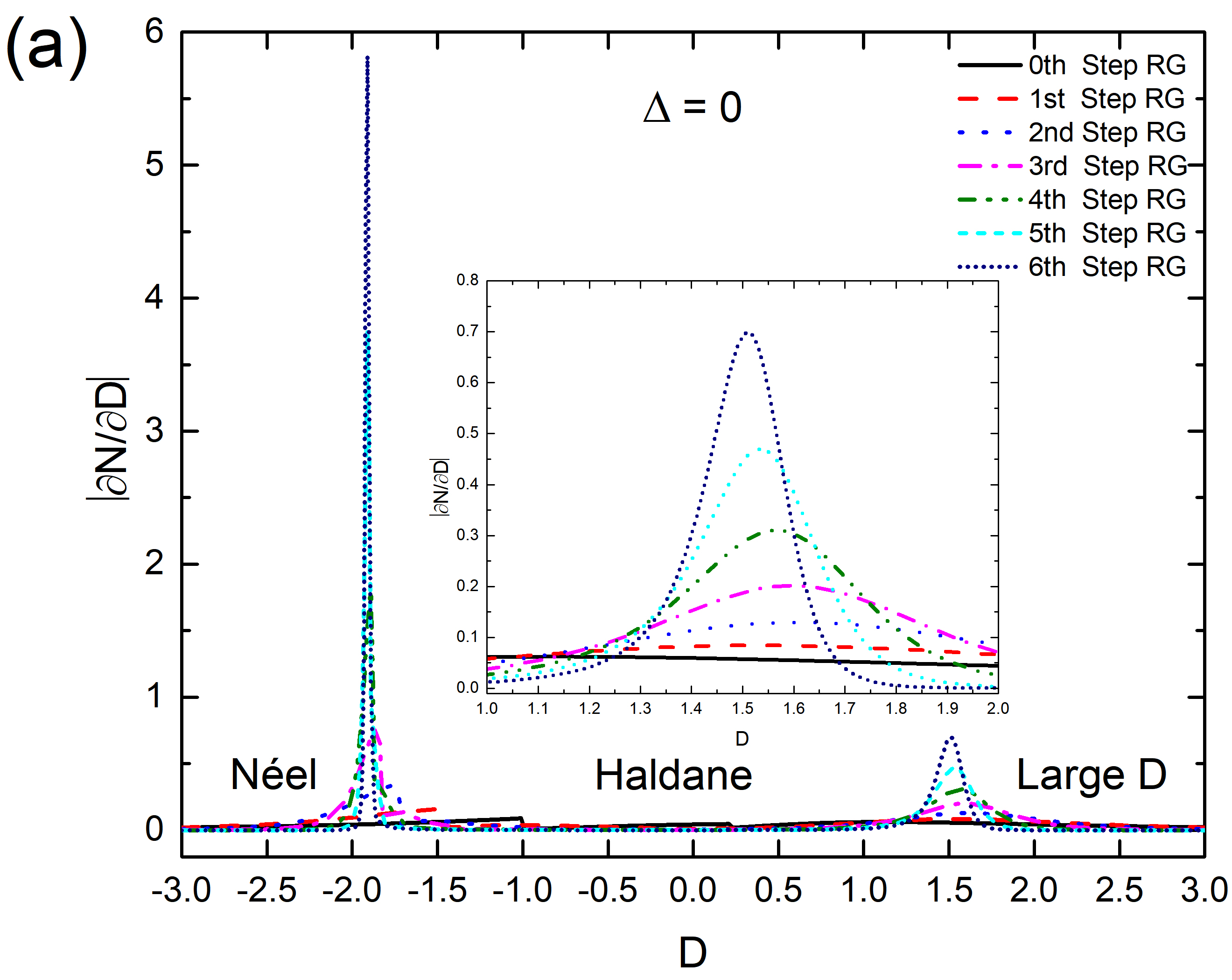}\includegraphics[width=83mm]{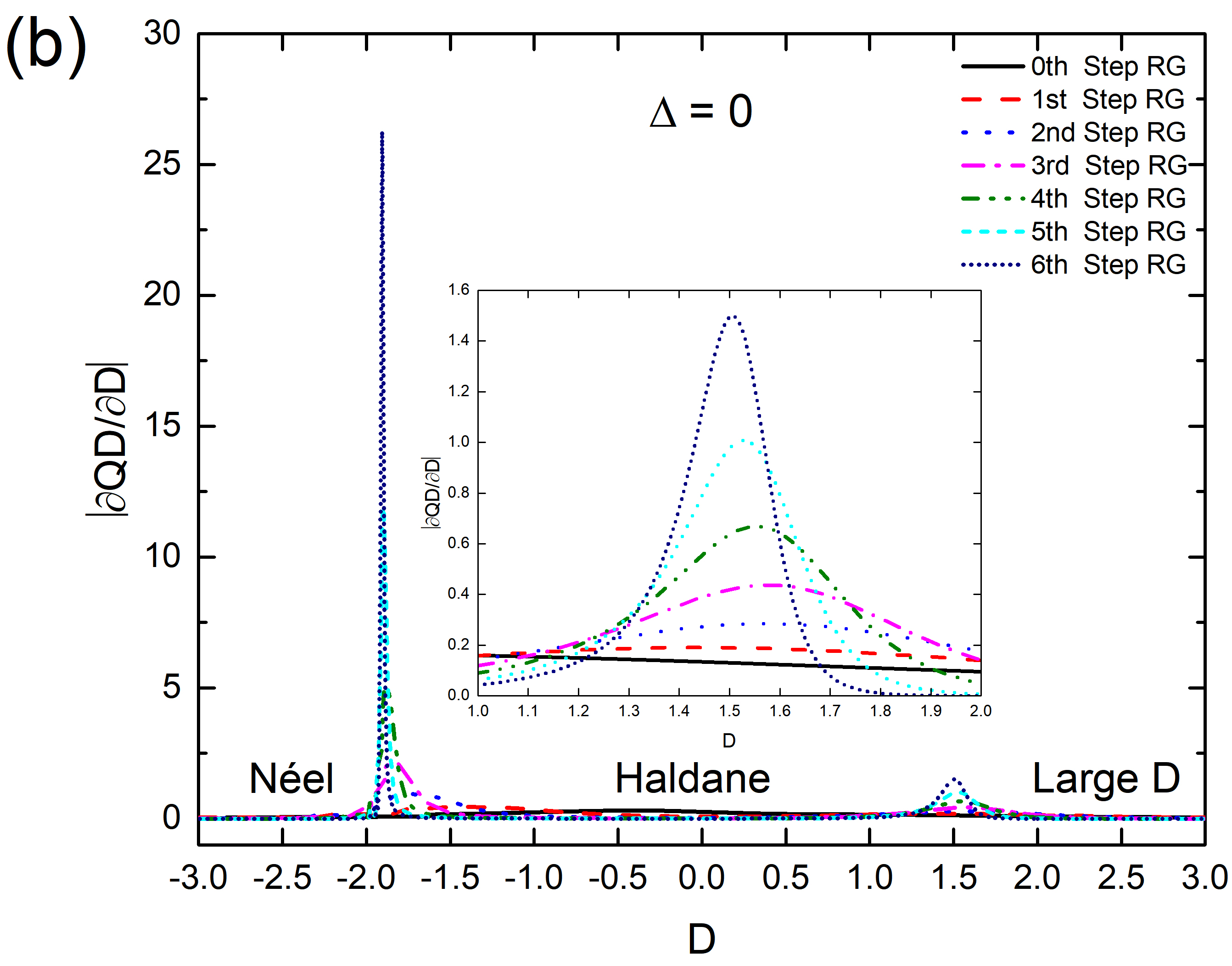}

\caption{Absolute value of the first partial derivative of (a) negativity and
(b) QD, with respect to $D$, as the step of the QRG iterations increases
at $\Delta=0$ (figure \ref{Figure 4}(a) and (b)). Each phase is
labelled by the black text.}
\label{Figure 7}
\end{figure}

\begin{figure}[ht]
\centering\includegraphics[width=82mm]{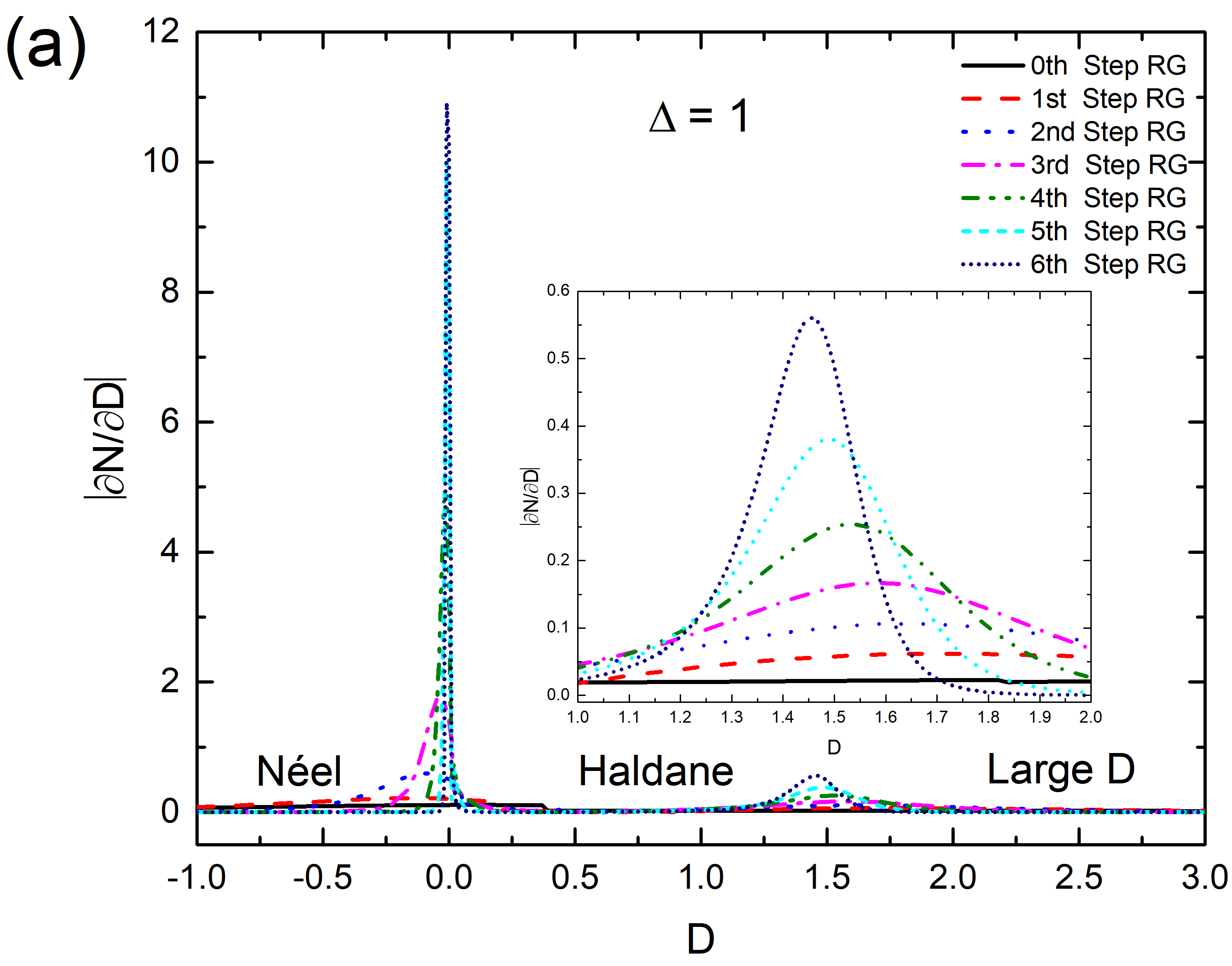}\includegraphics[width=82mm]{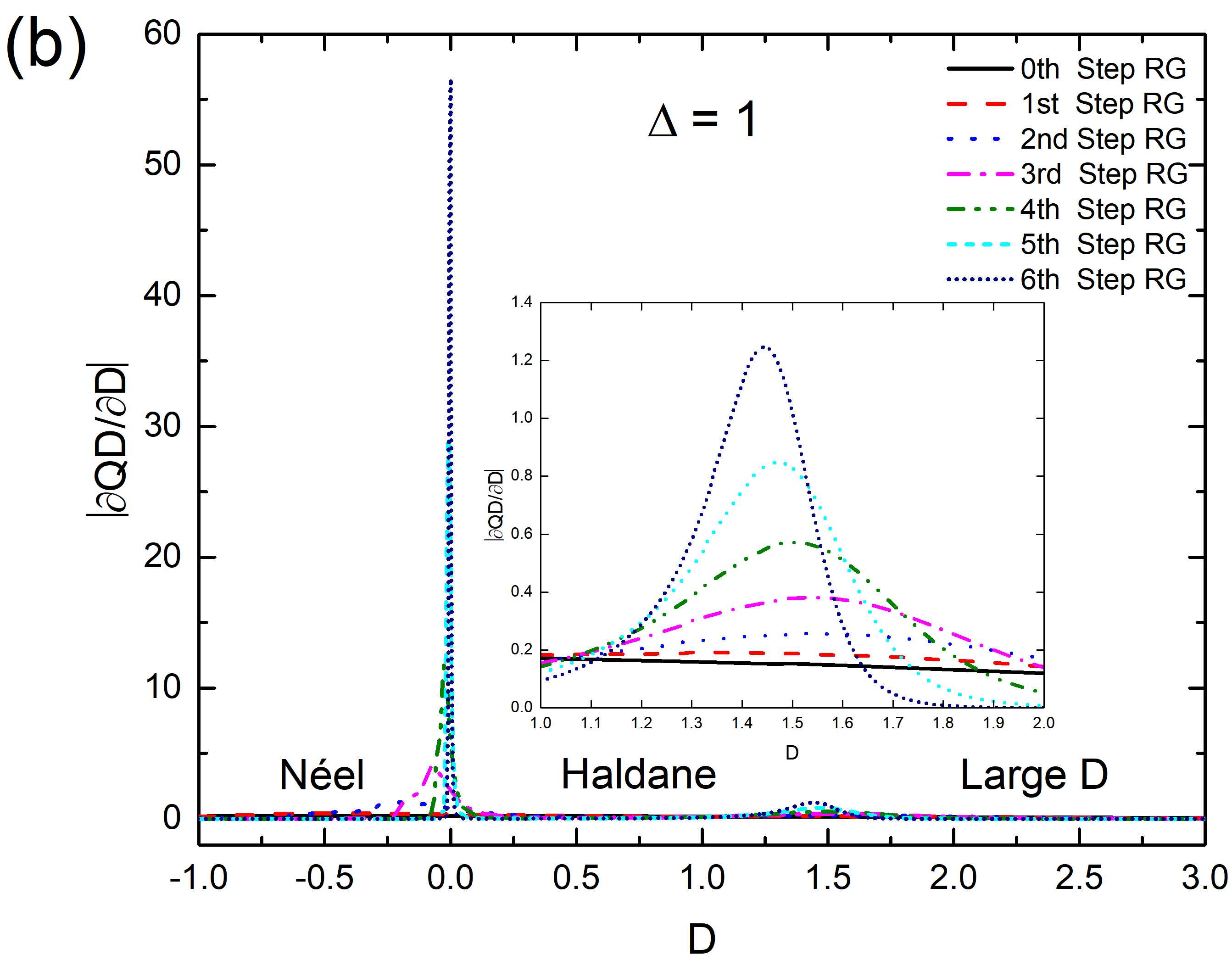}

\caption{Absolute value of the first partial derivative of (a) negativity and
(b) QD, with respect to $D$, as the step of the QRG iterations increases
at $\Delta=1$ (figure \ref{Figure 5}(a) and (b)). Each phase is
labelled by the black text.}
\label{Figure 8}
\end{figure}

\begin{figure}[ht]
\centering\includegraphics[width=82mm]{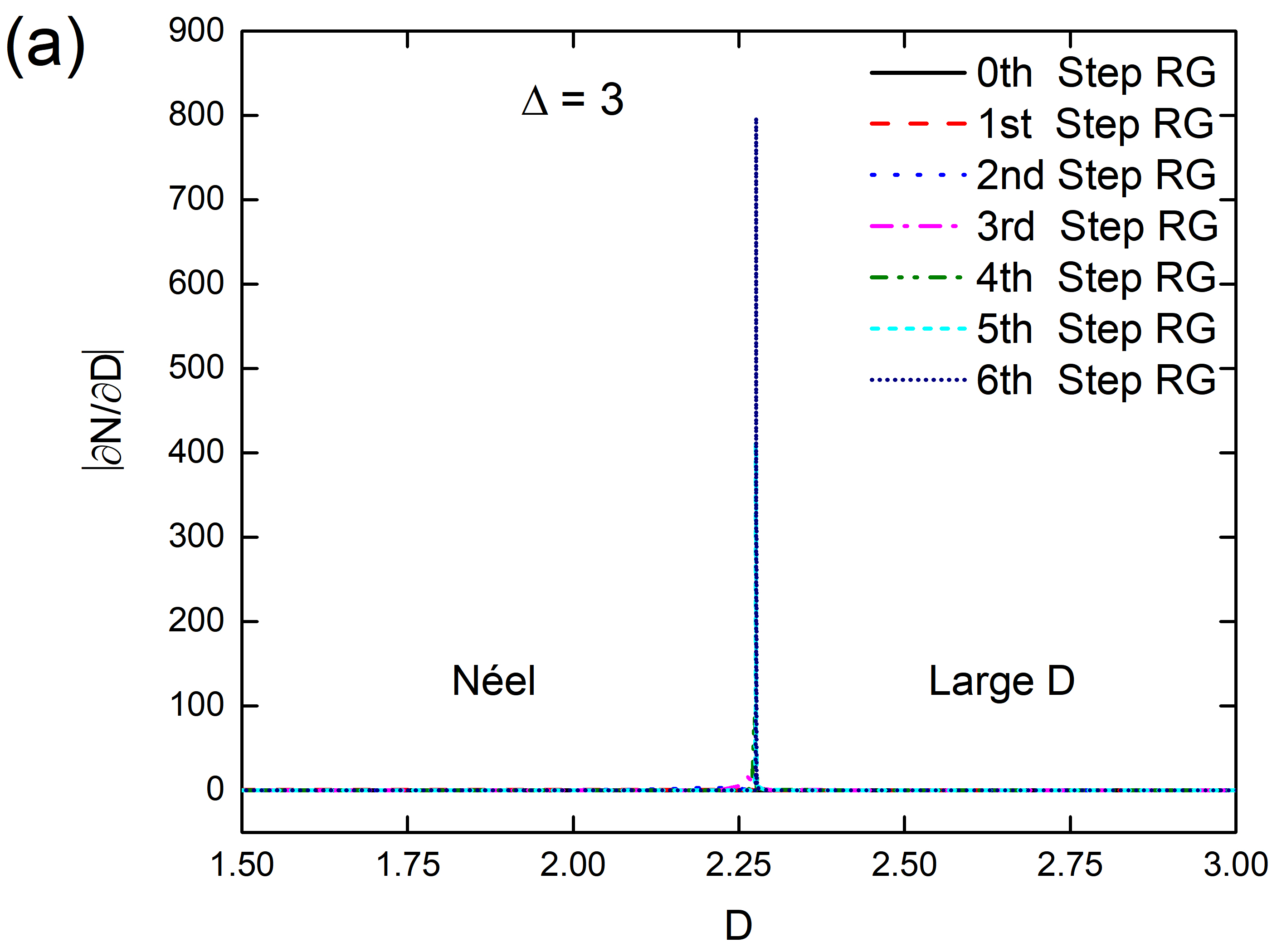}\includegraphics[width=82mm]{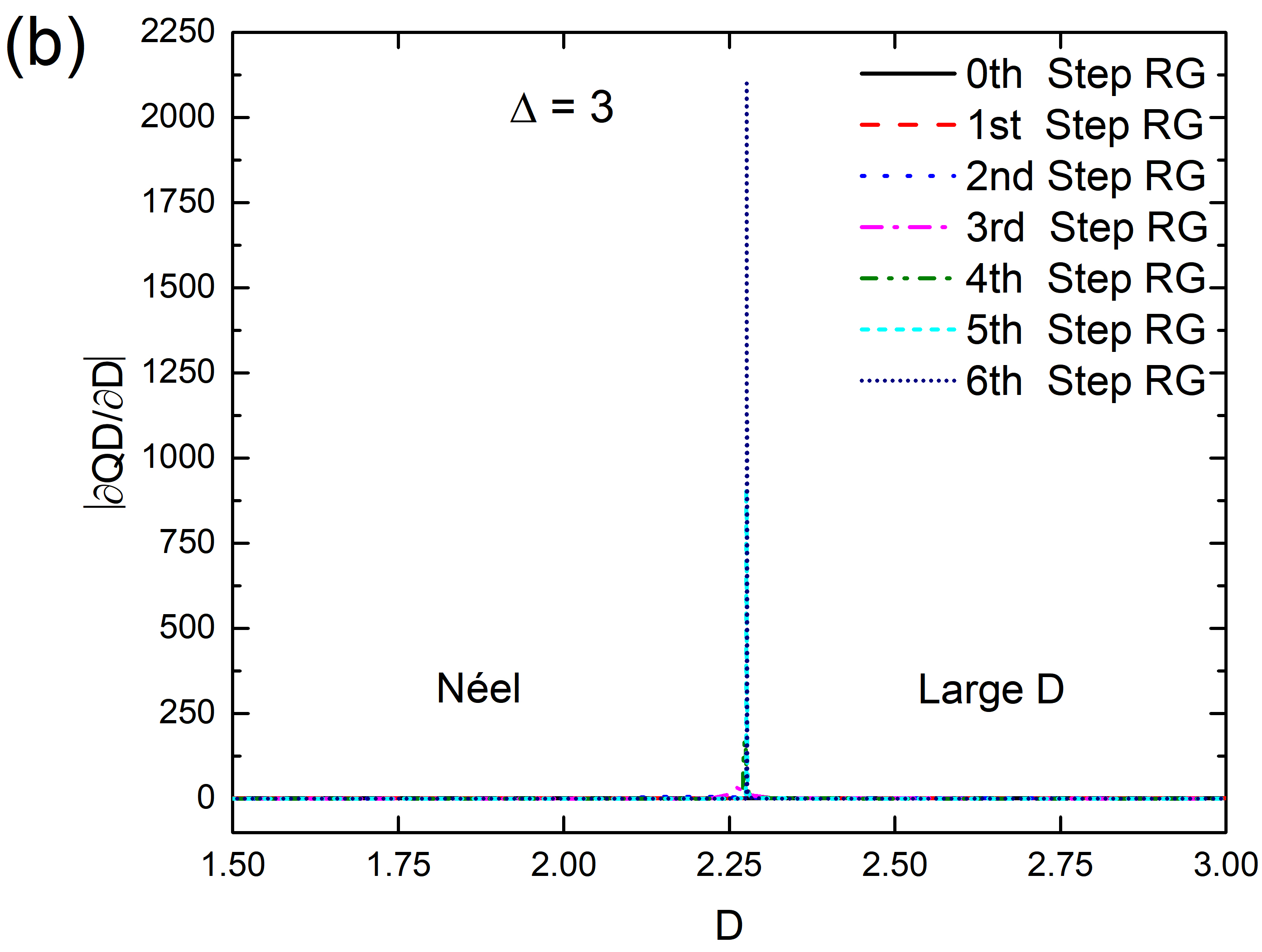}

\caption{Absolute value of the first partial derivative of (a) negativity and
(b) QD, with respect to $D$, as the step of the QRG iterations increases
at $\Delta=3$ (figure \ref{Figure 6}(a) and (b)). Each phase is
labelled by the black text.}
\label{Figure 9}
\end{figure}

The entanglement and QC usually present nonanalytic behaviors at the
phase transition points, which are accompanied by scaling behavior
due to the divergence of the correlation length. This section shows
the QPT and nonanalytic behaviors of the negativity and QD in the
spin-1 Heisenberg chain. We analyze the first partial derivative of
negativity and QD with respect to the single-ion anisotropy parameter
for a fixed value of $\bigtriangleup=0$. As shown in \ref{Figure 7}(a),
the absolute value of the first partial derivative of the negativity,
with respect to the single-ion anisotropy parameter, i.e., $\left\vert \partial N/\partial D\right\vert $,
is discontinuous at the critical point $P_{4}(0,-1.91)$, and the
singular behavior becomes more pronounced as the size of the system
increases. However there is only a maximum at the critical point $P_{3}:(0,1.45)$,
up to 6th step of RG, as shown in the inset of the same figure. The
single-ion anisotropy parameter corresponds to the maximum of $\left\vert \partial N/\partial D\right\vert $
tends to the fixed point as the size of the system increases. The
QD exhibits similar behavior with respect to the single-ion anisotropy
parameter, i.e., $\left\vert \partial QD/\partial D\right\vert $,
as shown in \ref{Figure 7}(b). For $\bigtriangleup=1$, $\left\vert \partial N/\partial D\right\vert $
shows a growing peak at the phase transition point $P_{1}:(1.0,0)$,
and the position of the extreme point varies as the size of the system
increases, which arises from finite-size effects. Note that, there
is only a maximum at $P_{6}:(1.0,1.37)$ in the 6th step of the RG,
as shown in figure \ref{Figure 8}(a). One can infer that, with enough
RG iterations, both $\left\vert \partial N/\partial D\right\vert $
and $\left\vert \partial QD/\partial D\right\vert $ should also exhibit
nonanalytic behavior at the boundary of Haldane and Large-D phases.
Similar behavior of the QD with respect to $D$ for $\bigtriangleup=1$
are also obtained, as shown in figure \ref{Figure 8}(b). For $\bigtriangleup=3$,
both $\left\vert \partial N/\partial D\right\vert $ and $\left\vert \partial QD/\partial D\right\vert $
are discontinuous at $P_{5}:(3.0,2.27)$, as shown in figure \ref{Figure 9}(a)
and (b). For $D=0$, both $\left\vert \partial N/\partial\bigtriangleup\right\vert $
and $\left\vert \partial QD/\partial\bigtriangleup\right\vert $ also
become discontinuous at $P_{1}:(1.0,0)$, as the size of the system
increases, as shown in figure S3(a) and (b) of appendix B.

\begin{figure}
\centering\includegraphics[width=82mm]{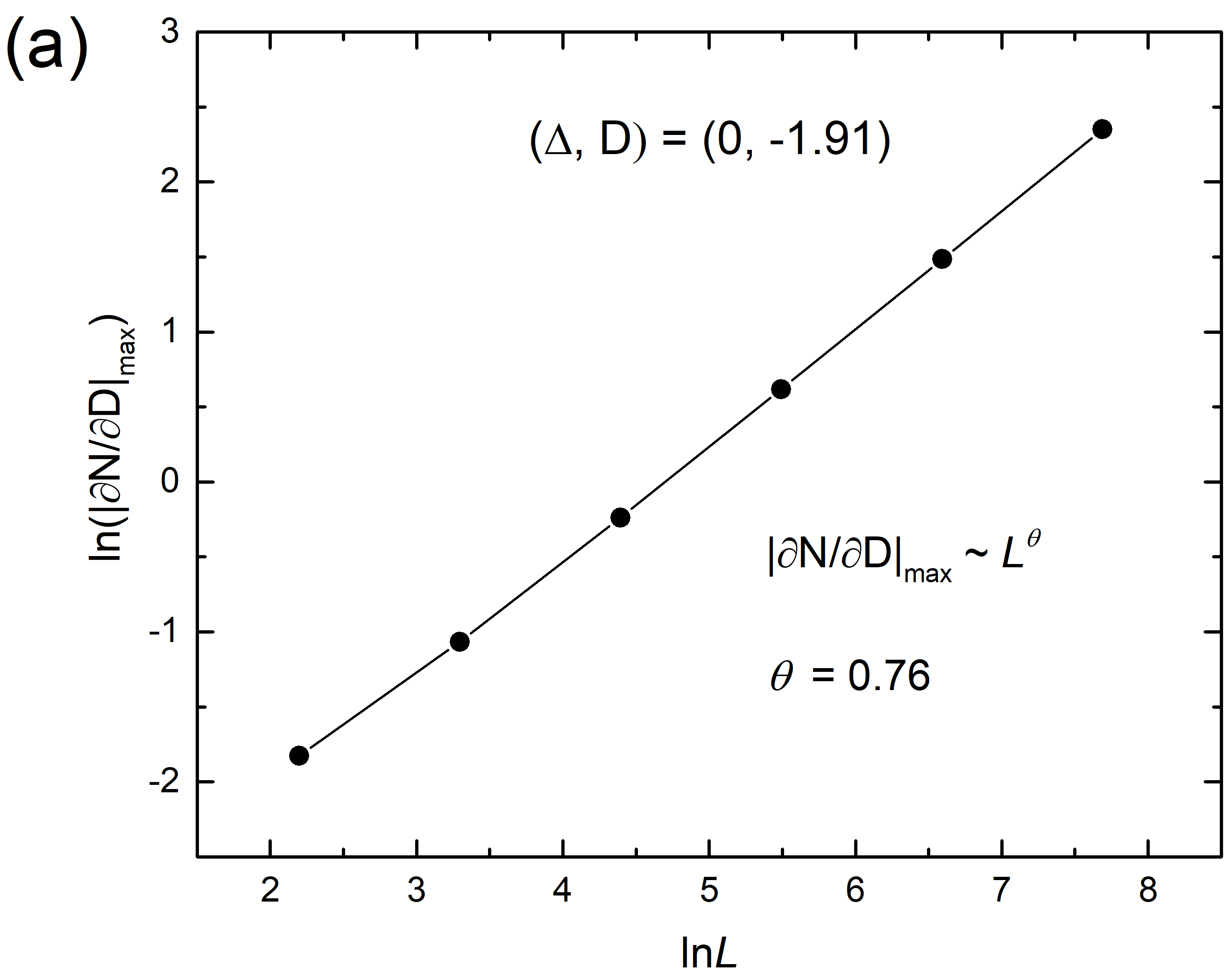}\includegraphics[width=82mm]{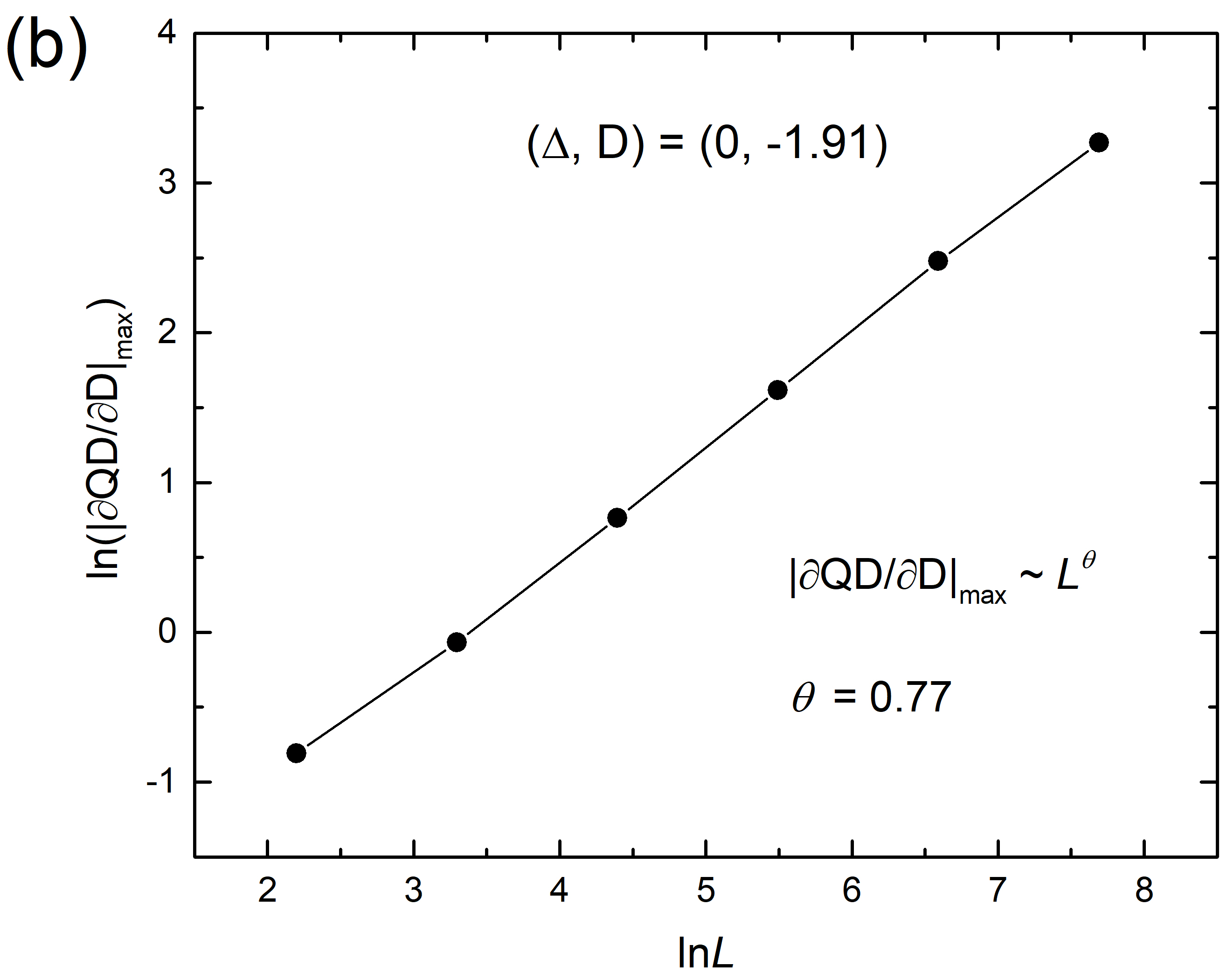}

:\includegraphics[width=82mm]{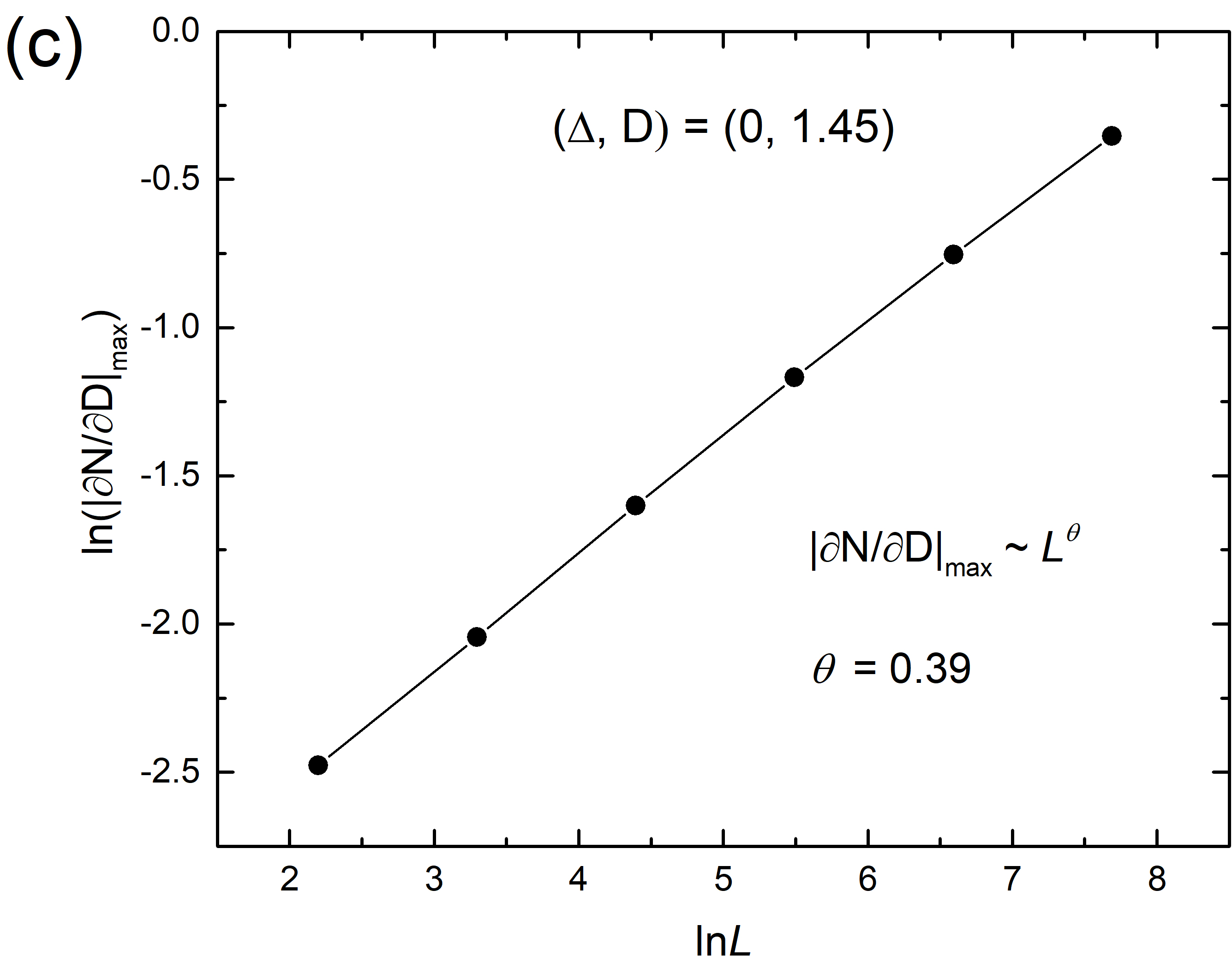}\includegraphics[width=82mm]{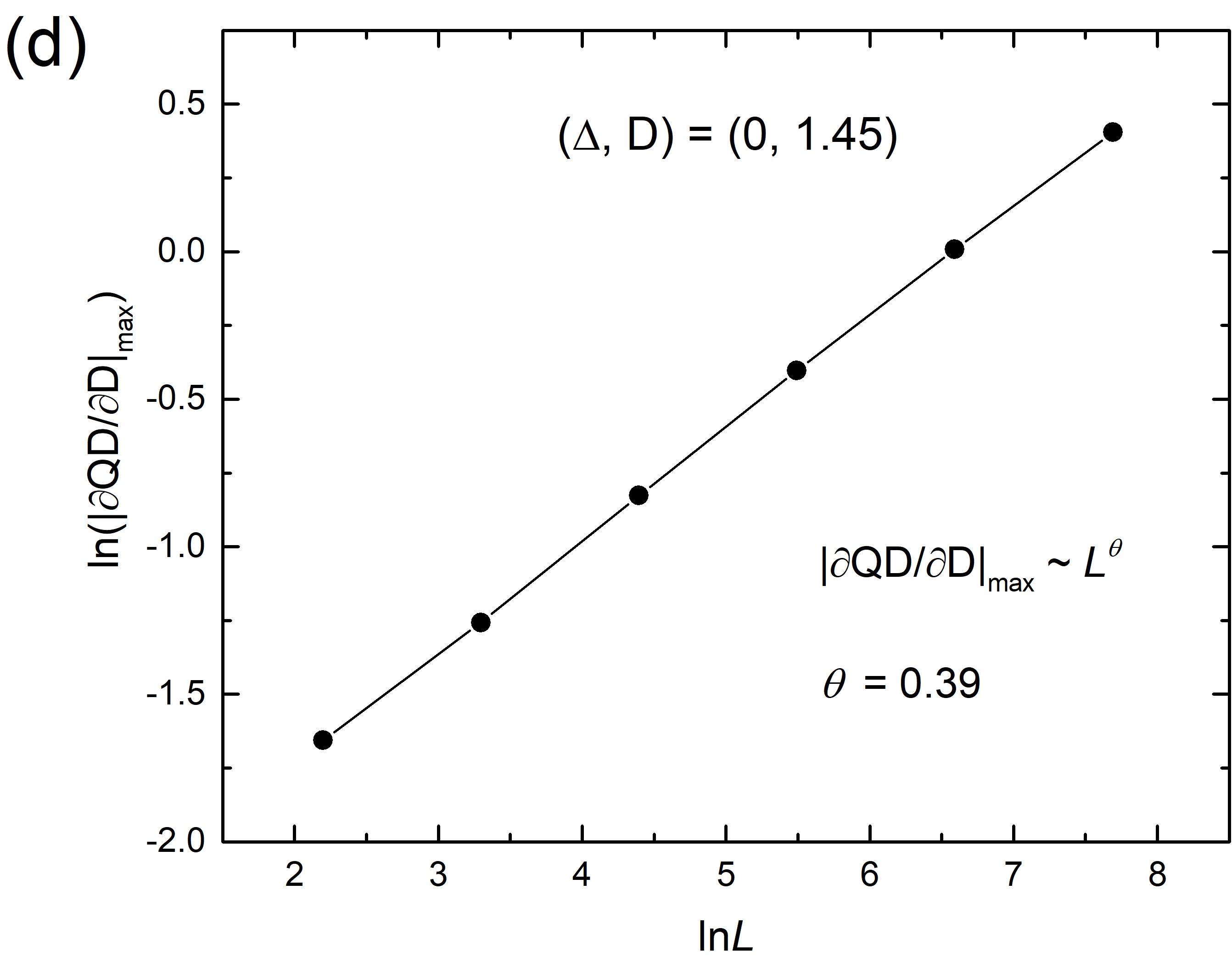}

\caption{Logarithm of the maximum of (a) $\left\vert \partial N/\partial D\right\vert $
and (b) $\left\vert \partial QD/\partial D\right\vert $ versus the
logarithm of the system size, $lnL$, at $\ensuremath{P_{4}(0,-1.91)}$.
Logarithm of the maximum of (c) $\left\vert \partial N/\partial D\right\vert $
and (d) $\left\vert \partial QD/\partial D\right\vert $ versus $lnL$
at $\ensuremath{P_{3}(0,1.45)}$.}
\label{Figure 10}
\end{figure}

The logarithm of the maximum of the absolute value of the first partial
derivative of the negativity, with respect to the single-ion anisotropy
$D$, i.e., $ln\left\vert \partial N/\partial D\right\vert _{max}$,
versus the logarithm of the system size, i.e., $lnL$, obey the linear
relation $\left\vert \partial N/\partial D\right\vert _{max}\sim L^{\theta}$
at the critical points $P_{4}(0,-1.91)$ and $P_{3}:(0,1.45)$. These
are shown in figure \ref{Figure 10}(a) and (c), respectively. The
singular behavior of the negativity and the scaling behavior of the
system depend on the QC exponent $\theta$, as shown in Table \ref{table1}.
A similar linear relation, i.e., $\left\vert \partial QD/\partial D\right\vert _{max}\sim L^{\theta}$
is also obtained at the critical points $P_{4}(0,-1.91)$ and $P_{3}:(0,1.45)$,
as shown in figure \ref{Figure 10}(b) and (d). The scaling behaviors
of $ln\left\vert \partial N/\partial D\right\vert _{max}$ versus
$lnL$ at $P_{1}:(1.0,0)$ and $P_{6}:(1.0,1.37)$ also exhibit linear
relations, as shown in figure S4(a) and figure S5(a) of appendix B,
respectively. Amazingly, for $D=0$, both $\left\vert \partial N/\partial\Delta\right\vert _{max}\sim L^{\theta}$
and $\left\vert \partial QD/\partial\Delta\right\vert _{max}\sim L^{\theta}$
have nearly the same relation at $P_{1}:(1.0,0)$, as shown in figure
S6(a) and (b). The linear relation of $\left\vert \partial N/\partial D\right\vert _{max}\sim L^{\theta}$
at $P_{5}:(3.0,2.27)$ is shown in figure S7(a) of appendix B. Furthermore,
the scaling behavior of $ln\left\vert \partial QD/\partial D\right\vert _{max}$
versus $lnL$ at the phase transition points $P_{1}:(1.0,0)$, $P_{6}:(1.0,1.37)$
and $P_{5}:(3.0,2.27)$ also exhibits the linear relations $\left\vert \partial QD/\partial D\right\vert _{max}\sim L^{\theta}$,
as shown in figure S4(b), figure S5(b), and figure S7(b) of appendix
B. The QC exponents $\theta$, calculated from the negativity and
QD, are nearly equal at each fixed point.

In the spin system, the correlation length $\xi$ becomes divergent
as $\xi\sim|D-D_{c}|^{-\nu}$ at the critical point $D_{c}$ and the
correlation length exponent $\nu$ can be obtained from the formula
$\nu=\left[ln\left(n_{B}\right)\right]/ln\left[\partial D^{^{\prime}}/\partial D\right]$$|_{D_{c}}$,
where $n_{B}$ is the number of spins in one block and $D^{^{\prime}}$
is the recurrence relation of the single-ion parameter $D$ in Eq.(\ref{4})
\cite{PhysRevB.78.214414,SIERRA}. In the Haldane-Néel transition
at $\ensuremath{P_{1}:(1.0,0)}$, the entanglement exponent $\theta=1.01$
is nearly equal to that of the Néel-Haldane transition, i.e. $\theta=0.97$,
which implies that both the Néel-Haldane and Haldane-Néel transitions
at $P_{1}:(1.0,0)$ are in the same universality class, even though
the correlation length exponent in the Haldane-Néel transition, $\nu=1.39$,
is much smaller than that of the Néel-Haldane transition $\nu=3.12$.
We can infer that, for a given fixed point, the phase transitions
along different directions are in the same universality class. The
type of phase transition, the QC exponent $\theta$, and the correlation
length exponent $\nu$ at each fixed point are presented in Table
\ref{table1}. Amazingly, both at the critical points $P_{3}:(0,1.45)$
and $P_{4}:(0,-1.91)$, their QC exponent $\theta$ and the corresponding
correlation length exponent $\nu$, are reciprocals, i.e., $\nu=\frac{1}{\theta}$.
This is similar to the case of spin-1/2 systems \cite{PhysRevA.83.062309,PhysRevA.84.042302}.
As the critical points are approaching the large-size limit (not the
thermodynamic limit), the correlation length covers the whole system.

\begin{table}
\caption{\label{table1}The type of phase transition, QC exponent $\theta$,
and the correlation length exponent $\nu$ of the spin-1 Heisenberg
model at each fixed point $\left(\bigtriangleup,D\right)$. The subscripts
$N$ and $QD$ indicate the results derived from negativity and QD,
respectively.}
\[
\begin{tabular}{|c|c|c|c|c|}
\hline  \ensuremath{(\bigtriangleup,D)}  &  transition type  &  \ensuremath{\theta_{N}}  &  \ensuremath{\theta_{QC}}  &  \ensuremath{\nu} \\
\hline  \ensuremath{P_{1}:(1.0,0)}  &  Néel-Haldane (Ising)  &  \ensuremath{0.97}  &  \ensuremath{0.97}  &  \ensuremath{3.12} \\
 \ensuremath{P_{3}:(0,1.45)}  &  Haldane-large \textit{D} (Gaussian)  &  \ensuremath{0.39}  &  \ensuremath{0.39}  &  \ensuremath{3.07} \\
 \ensuremath{P_{4}(0,-1.91)}  &  Néel-Haldane (Ising)  &  \ensuremath{0.76}  &  \ensuremath{0.77}  &  \ensuremath{1.26} \\
 \ensuremath{P_{5}:(3.0,2.27)}  &  Néel-large \textit{D} (first-order)  &  \ensuremath{\ensuremath{1.52}}  &  \ensuremath{\ensuremath{1.57}}  &  \ensuremath{1.58} \\
 \ensuremath{P_{6}:(1.0,1.37)}  &  Haldane-large \textit{D} (Gaussian)  &  \ensuremath{0.4}  &  \ensuremath{0.35}  &  \ensuremath{\ensuremath{5.69}}
\\\hline \end{tabular}
\]
\end{table}

\section{Summary}

The QCs and QPTs of the spin-1 Heisenberg chain with single-ion anisotropy
were investigated using the QRG method. The phase diagram of the spin-1
system is more complex and richer than that of the spin-1/2 system,
which is determined by the easy-axis anisotropy and single-ion anisotropy
parameters. Both negativity and QD can equivalent to depict the QPT.
The single-ion anisotropy parameter plays an important role in reducing
the negativity and QD by favoring the alignment of spins. As the scale
of the system increases, the negativity and QD exhibit step-like patterns
in different phases. The critical behavior of the spin-1 chain was
described by the first partial derivative of the negativity or QD
of the blocks, which show nonanalytic behavior at the phase transition
points. The QC exponent $\theta$ and correlation length exponent
$\nu$ derived from negativity and QD are nearly equal at each fixed
point. Notably, they are reciprocals, i.e. $\nu=\frac{1}{\theta}$,
at the critical points $P_{3}:(0,1.45)$ and $P_{4}:(0,-1.91)$. This
is similar to the case of spin-1/2 systems \cite{PhysRevB.78.214414,PhysRevA.83.062309,PhysRevA.84.042302}
and our results are also consistent with previous work \cite{Langari_2013}.
Better yet, the entanglement and QD show a clear QPT, even when the
scale of spin-1 systems is as small as $3^{3}$ sites.
\begin{acknowledgments}
One of the authors, Wanxing Lin, would like to thank Dao-Xin Yao and  Shi-Dong
Liang for their encouragements. He also thanks Bao-Tian Wang, Jun-Qing
Cheng, Matthew J Lake, A Langari, and M Siahatgar for stimulating
discussions. This work is supported by the National Natural Science
Foundation of China (nos. 11675090, 11905095, 11847086, 11505103,
and 11275112), the Shandong Natural Science Foundation (nos. ZR2019PA015
and ZR2011AM018), and the Specialized Research Fund for the Doctoral
Program of Higher Education (no. 20123705110004).
\end{acknowledgments}

\clearpage{}

\appendix
%dummy comment inserted by tex2lyx to ensure that this paragraph is not empty%dummy comment inserted by tex2lyx to ensure that this paragraph is not empty%dummy comment inserted by tex2lyx to ensure that this paragraph is not empty%dummy comment inserted by tex2lyx to ensure that this paragraph is not empty%dummy comment inserted by tex2lyx to ensure that this paragraph is not empty%dummy comment inserted by tex2lyx to ensure that this paragraph is not empty%dummy comment inserted by tex2lyx to ensure that this paragraph is not empty

\section{The QRG procedure}

The quantum renormalization group procedure can be reedited as shown
below \cite{Langari_2013}:

To begin with, the lattice is decomposed into isolated blocks where
the total Hamiltonian is written as a sum of isolated block Hamiltonians
$\left(H^{B}\right)$ and inter-block interactions $\left(H^{BB}\right)$,
i.e., $H=H^{B}+H^{BB}$, where, $H^{B}=\sum\nolimits _{I=1}^{L/3}h_{I}^{B}$,
$H^{BB}=\sum\nolimits _{I=1}^{L/3}h_{I,I+1}^{BB}$, and
\[
h_{I}^{B}=J\left[\sum_{j=1}^{2}\left(S_{I,j}^{x}S_{I,j+1}^{x}+S_{I,j}^{y}S_{I,j+1}^{y}+\bigtriangleup S_{I,j}^{z}S_{I,j+1}^{z}\right)+D\sum_{j=1}^{3}\left(S_{I,j}^{z}\right)^{2}\right],\tag{a1}
\]
\[
h_{I,I+1}^{BB}=J\left(S_{I,3}^{x}S_{I+1,1}^{x}+S_{I,3}^{y}S_{I+1,1}^{y}+\bigtriangleup S_{I,3}^{z}S_{I+1,1}^{z}\right).\tag{a2}
\]
$S_{I,j}^{\alpha}$ denotes the $\alpha$-component of the $j$th
spin in block $I$. The energy eigenstates of $h_{I}^{B}$ are calculated
exactly and the three lowest eigenvectors are denoted by $\left\vert \phi_{0}\right\rangle $
and $\left\vert \phi_{\pm}\right\rangle $ with the corresponding
eigenvalues $E_{0}$ and $E_{1}$, respectively.

In addition, the three low-lying energy eigenstates of each block
are kept to build up an embedding (projection) operator $\left(T\right)$,
representing the most important subspace of the original Hilbert space
$\left(H\right)$. The embedding operator for each block is constructed
as
\[
T_{I}=\left\vert \phi_{+}\right\rangle \left\langle +1\right\vert +\left\vert \phi_{0}\right\rangle \left\langle 0\right\vert +\left\vert \phi_{-}\right\rangle \left\langle -1\right\vert ,\tag{a3}
\]
where $\left\vert \pm1\right\rangle ,\left\vert 0\right\rangle $
are the base kets for the renormalized Hilbert space of each block.

Finally, the original Hamiltonian $\left(H\right)$ is mapped into
the renormalized Hamiltonian $\left(H^{^{\prime}}\right)$ utilizing
the\textbf{\ }embedding operator, which is given by
\[
H^{^{\prime}}=\sum_{I=1}^{L/3}\left(T_{I}^{\dagger}h_{I}^{B}T_{I}+T_{I}^{\dagger}T_{I+1}^{\dagger}h_{I,I+1}^{BB}T_{I+1}T_{I}\right).\tag{a4}
\]
The first part of the projections leads to
\[
T_{I}^{\dagger}h_{I}^{B}T_{I}=E_{0}1+\left(E_{1}-E_{0}\right)\left(S_{I}^{z}\right)^{2},\tag{a5}
\]
and the second term of the projection defines the effective interaction
between blocks $I$ and $I+1$ in terms of the renormalized operators,
\[
T_{I}^{\dagger}S_{I,j}^{\alpha}T_{I}=X_{ren}S_{I}^{^{\prime\alpha}};~j=1,3;~\alpha=x,y,\tag{a6}
\]
\[
T_{I}^{\dagger}S_{I,j}^{z}T_{I}=Z_{ren}S_{I}^{^{\prime z}};~j=1,3.\tag{a7}
\]

The renormalization coefficients $X_{ren}$ and $Z_{ren}$ are given
by the following expressions,

\begin{align}
X_{ren} & =\frac{1}{\sqrt{A_{5}A_{9}}}[2\left(E_{0}-2D\right)+2\left(E_{1}-3D\right)\nonumber \\
 & \times\left[4D^{2}-2D\left(\bigtriangleup+2E_{0}\right)+E_{0}\left(\bigtriangleup+E_{0}\right)-2\right]\nonumber \\
 & -A_{2}A_{3}\left(D-E_{1}\right)[4D^{2}E_{0}-2A_{8}-2D\nonumber \\
 & \times\left[E_{0}\left(\bigtriangleup+2E_{0}\right)-3\right]+E_{0}^{2}\left(\bigtriangleup+E_{0}\right)-6E_{0}]\nonumber \\
 & +2A_{1}A_{2}A_{3}\left(A_{8}-D+E_{0}\right)\nonumber \\
 & -\frac{A_{7}\left[A_{1}A_{2}A_{3}\left(D-E_{1}\right)-6D+2E_{1}\right]}{2D-E_{0}}],\tag{a8}
\end{align}
\begin{align}
Z_{ren} & =\frac{1}{A_{5}}[-[\left(D-E_{1}\right)[A_{1}A_{2}A_{3}\left(D-E_{1}\right)+4E_{1}-12D]+2]^{2}\nonumber \\
 & +4\left(E_{1}-3D\right)^{2}+4+A_{3}^{2}A_{2}^{2}\left(D-E_{1}\right)^{2}].\tag{a9}
\end{align}

The constants $A_{i}$ are given by:
\begin{align}
A_{1}=2\bigtriangleup-3D+E_{1},\tag{a10}
\end{align}
\begin{align}
A_{2}=E_{1}^{2}-4DE_{1}+3D^{2}-1,\tag{a11}
\end{align}
\begin{align}
A_{3}=\frac{1}{\bigtriangleup-2D+E_{1}},\tag{a12}
\end{align}
\begin{align}
A_{4}=\frac{1}{\bigtriangleup-2D+E_{0}},\tag{a13}
\end{align}
\begin{align}
A_{5} & =A_{3}^{2}\left[A_{1}A_{2}\left(D-E_{1}\right)+2\left(3D-E_{1}\right)\left(2D-\bigtriangleup-E_{2}\right)\right]^{2}\nonumber \\
 & +A_{2}^{2}A_{3}^{2}\left(D-E_{1}\right)^{2}+A_{1}^{2}A_{2}^{2}A_{3}^{2}+4\left(E_{1}-3D\right)^{2}+4\nonumber \\
 & +\left[\left(D-E_{1}\right)\left[A_{1}A_{2}A_{3}\left(D-E_{1}\right)+4\left(E_{1}-3D\right)\right]+2\right]^{2},\tag{a14}
\end{align}
\begin{align}
A_{6} & =16D^{4}E_{0}-8D^{3}\left[2E_{0}\left(\bigtriangleup+2E_{0}\right)-3\right]\nonumber \\
 & +4D^{2}\left[6E_{0}^{3}+6\bigtriangleup E_{0}^{2}+\left(\bigtriangleup^{2}-12\right)E_{0}-3\bigtriangleup\right]\nonumber \\
 & -2D\left[4E_{0}^{4}+6\bigtriangleup E_{0}^{3}+\left(2\bigtriangleup^{2}-15\right)E_{0}^{2}-9\bigtriangleup E_{0}+4\right]\nonumber \\
 & +E_{0}^{2}\left(\bigtriangleup+E_{0}\right)\left[E_{0}\left(\bigtriangleup+E_{0}\right)-6\right]+6E_{0},\tag{a15}
\end{align}
\begin{align}
A_{7}=-E_{0}^{3}+E_{0}^{2}\left(4D-\bigtriangleup\right)+2E_{0}\left(D\bigtriangleup-2D^{2}\right)-4D+2,\tag{a16}
\end{align}
\begin{align}
A_{8}=\frac{A_{4}\left(3E_{0}-4D\right)}{2D-E_{0}},\tag{a17}
\end{align}
\begin{align}
A_{9} & =\frac{A_{4}^{2}A_{6}^{2}+A_{7}^{2}}{\left(E_{0}-2D\right)^{2}}+2\left(E_{0}-2D\right)^{2}+4\left(A_{8}-D+E_{0}\right)^{2}+4\nonumber \\
 & +\left[4D^{2}-2D\left(\bigtriangleup+2E_{0}\right)+E_{0}\left(\bigtriangleup+E_{0}\right)-2\right]^{2}.\tag{a18}
\end{align}
The renormalized coupling constants can be obtained as shown in Eqs.(\ref{2})(\ref{3})(\ref{4})
of the text.

\clearpage{}

\section{The Figures}
See figures S1, S2, S3, S4, S5, S6 and S7.
\begin{figure}[ht]
\centering\includegraphics[width=82mm]{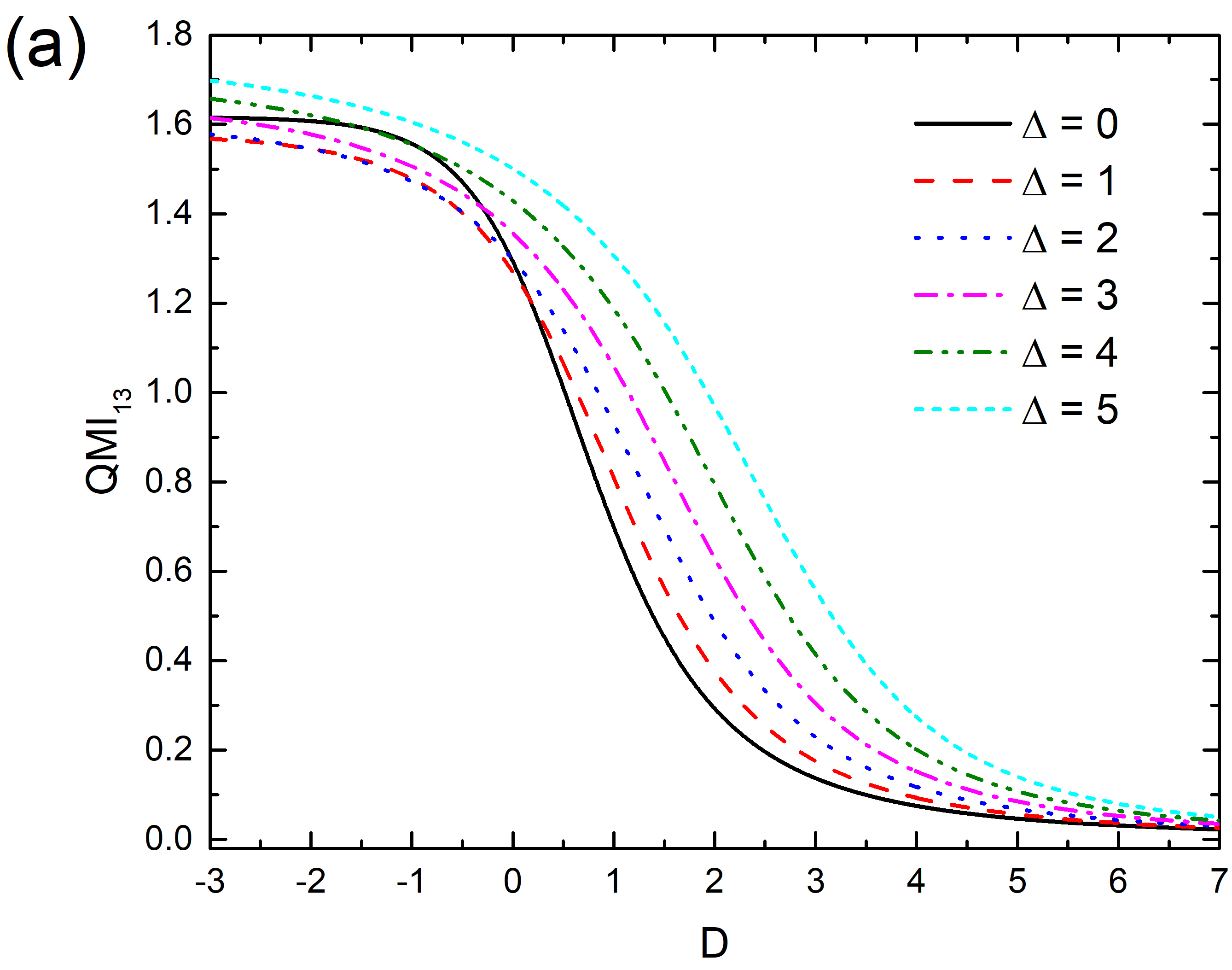}\includegraphics[width=82mm]{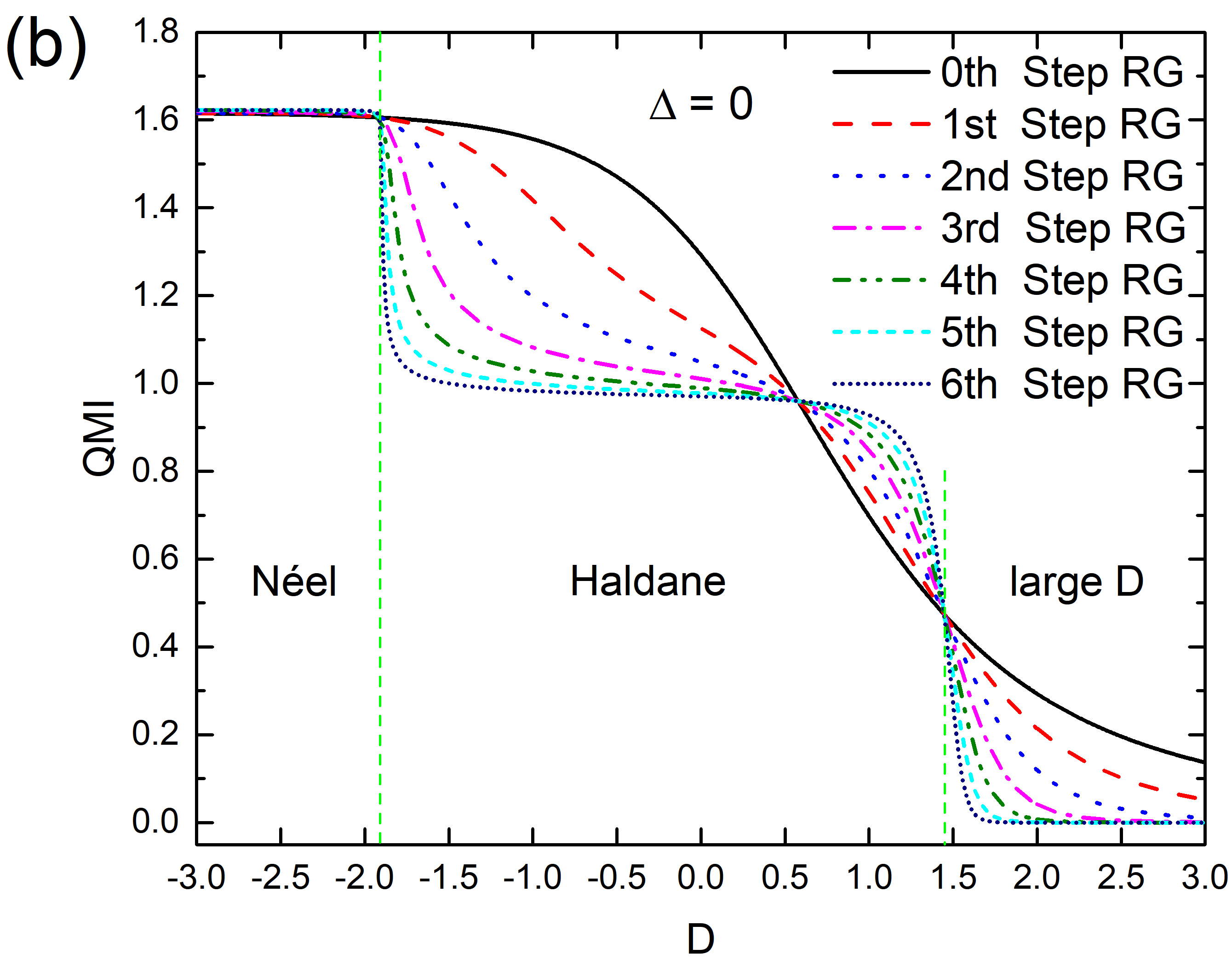}

\justifying{Figure S1. (a) The quantum mutual information (QMI)
between the first and third sites of the three-site model in terms
of the single-ion anisotropy $D$ for different $\Delta$. (b) The
quantum mutual information in terms of the QRG iterations at $\Delta=0$.
Each phase is marked by the black text, and separated by the dashed
green lines at $D=-1.91$ and $D=1.45$, respectively.}
\end{figure}

\begin{figure}[ht]
\centering\includegraphics[width=82mm]{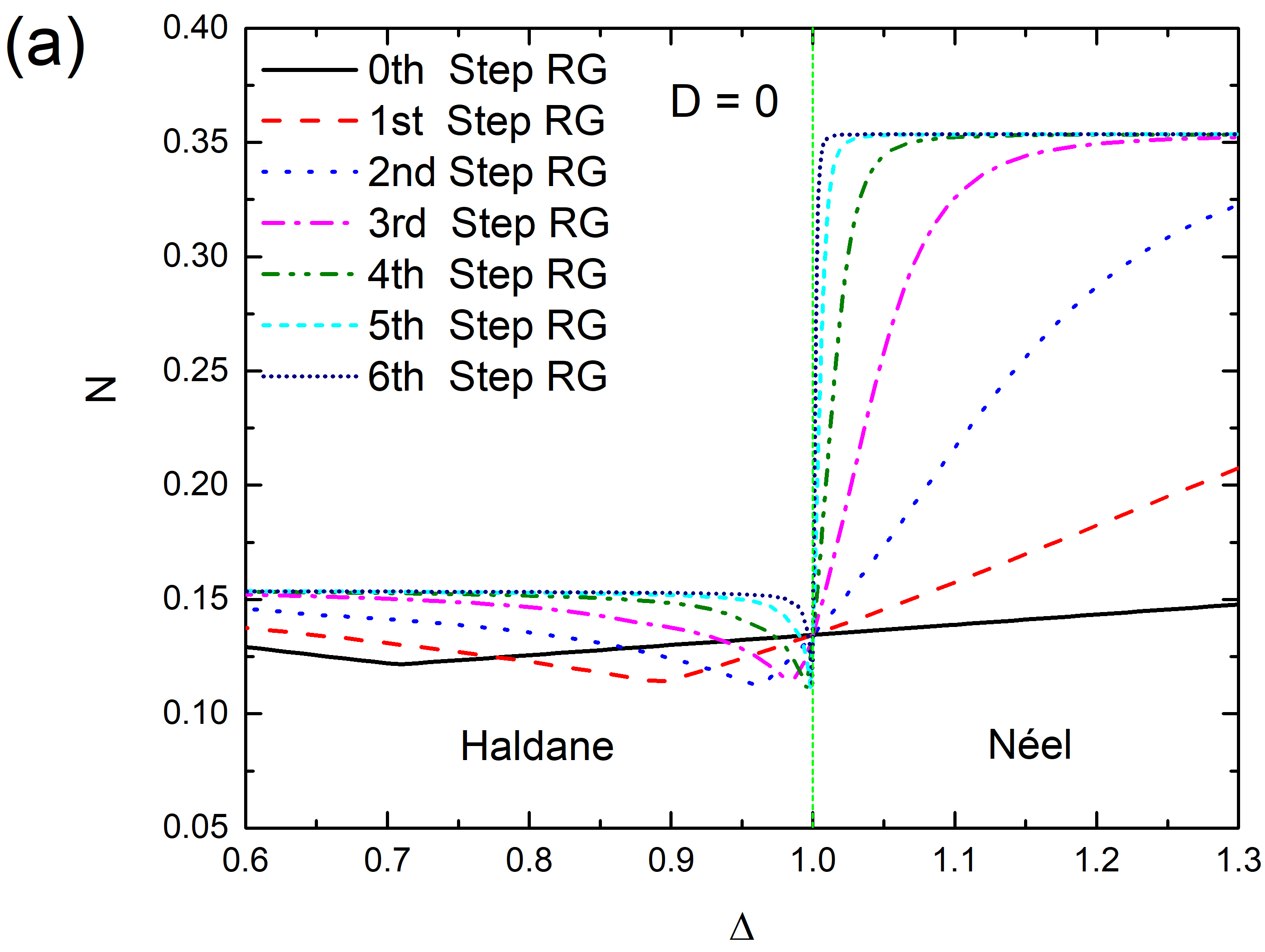}\includegraphics[width=81mm]{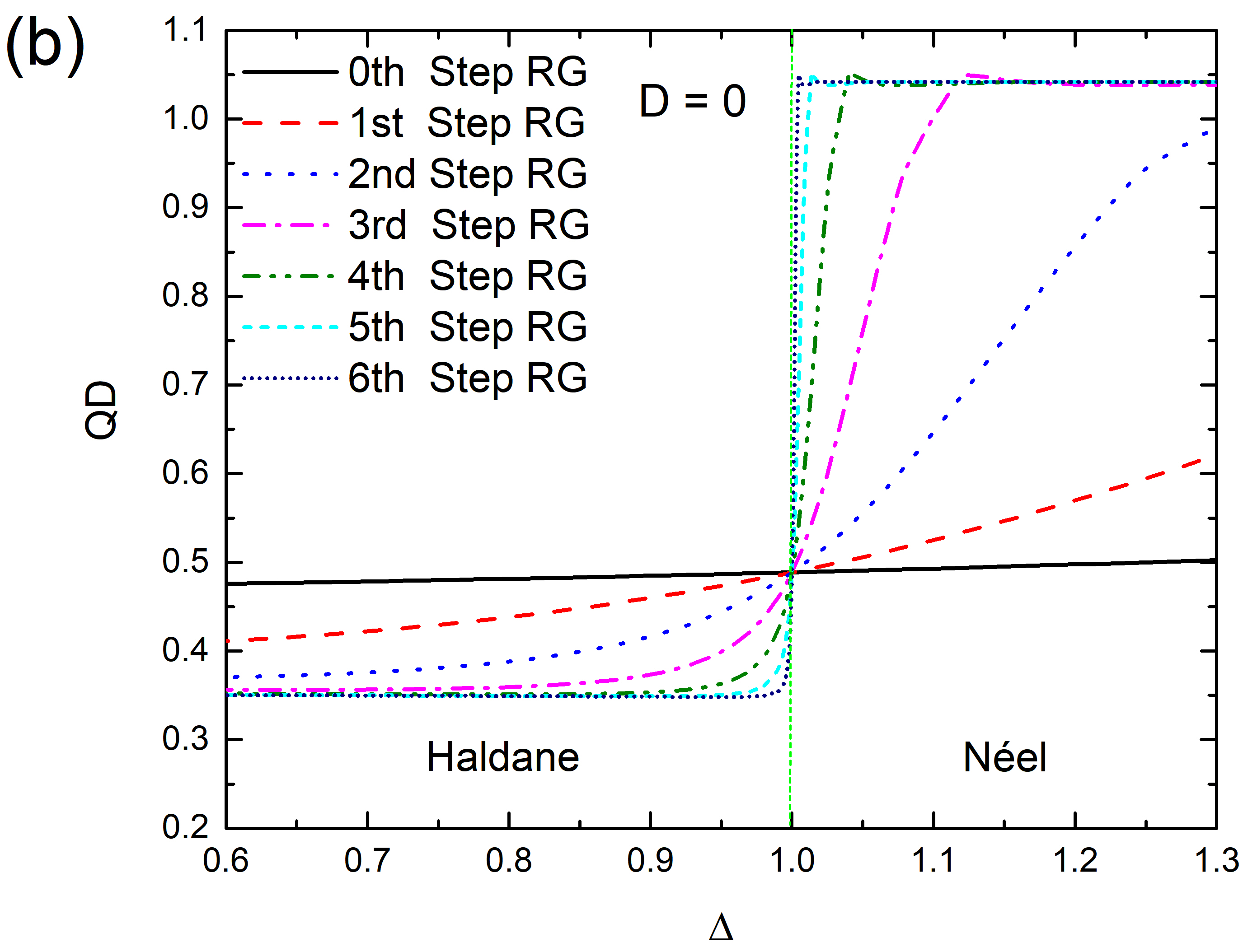}

\justifying{Figure S2. (a) Negativity and (b) QD in terms of the
QRG iterations at $D=0$. Each phase is marked by the black text,
and separated by the dashed green line at $\Delta=1.0$.}
\end{figure}

\begin{figure}[ht]
\centering\includegraphics[width=81mm]{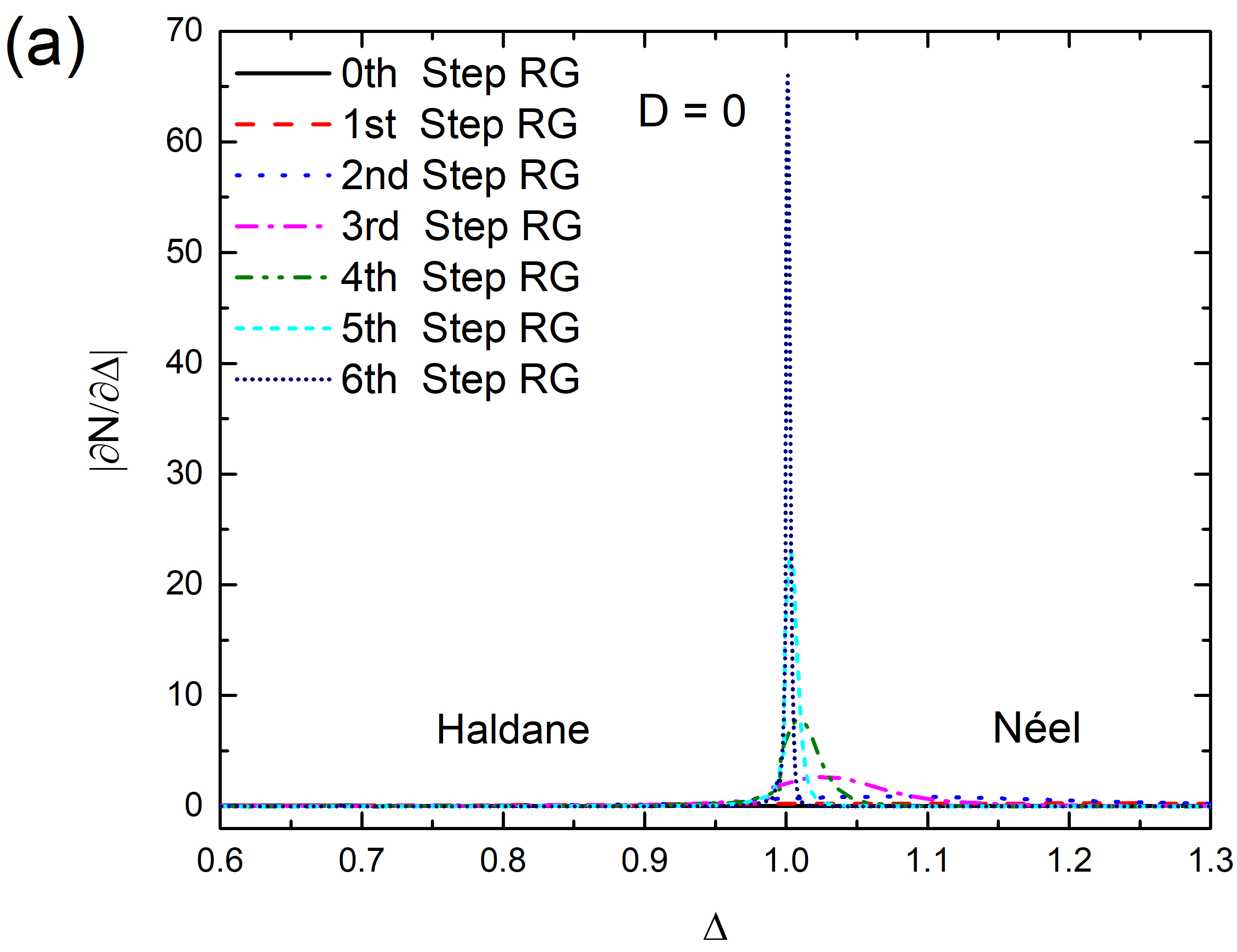}\includegraphics[width=82mm]{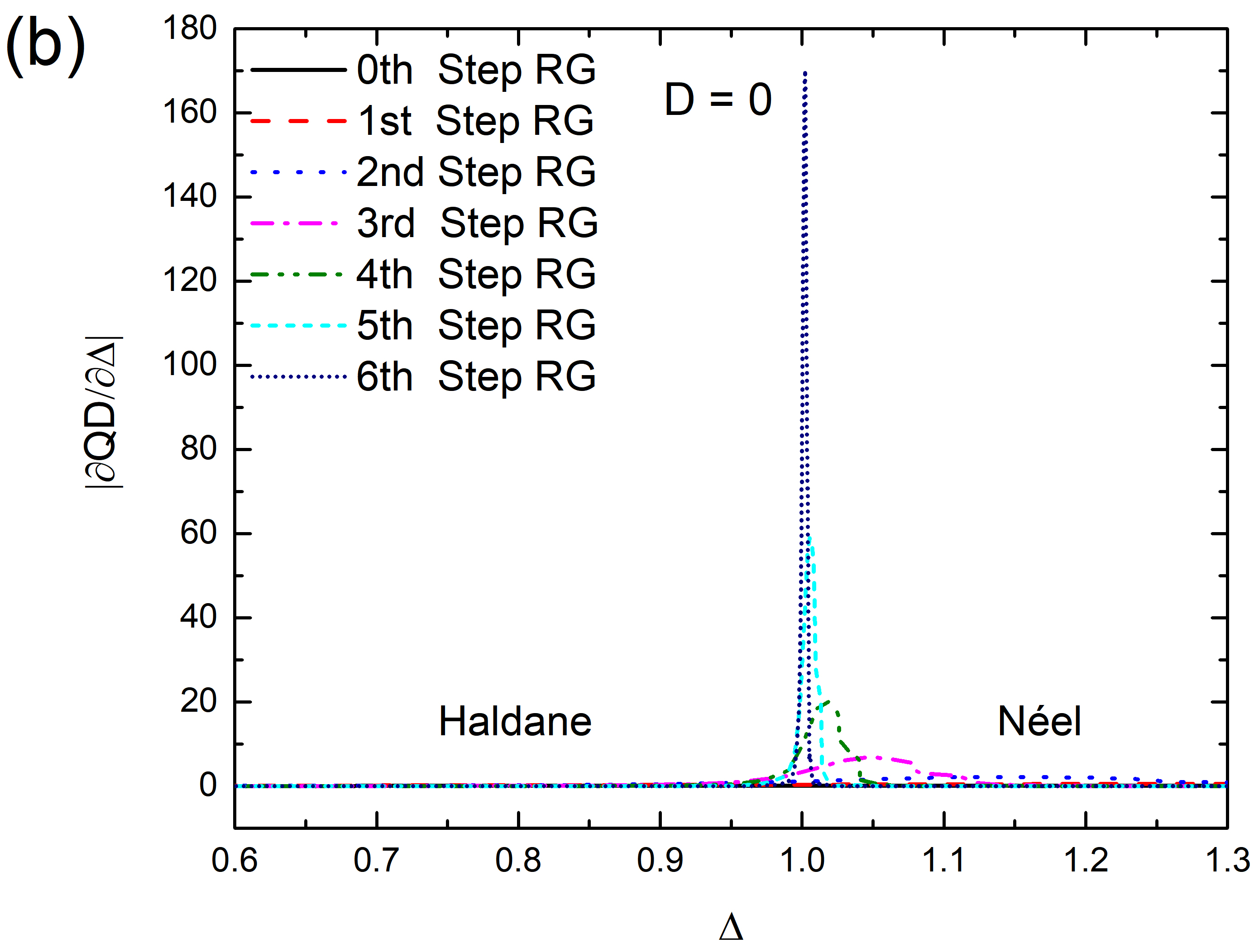}

\justifying{Figure S3. Absolute value of the first partial derivative
of (a) negativity and (b) QD, with respect to $\Delta$, as the step
of the QRG iterations increases at $D=0$ (figure S2(a) and (b)).
Each phase is labelled by the black text.}
\end{figure}

\begin{figure}[ht]
\centering\includegraphics[width=82mm]{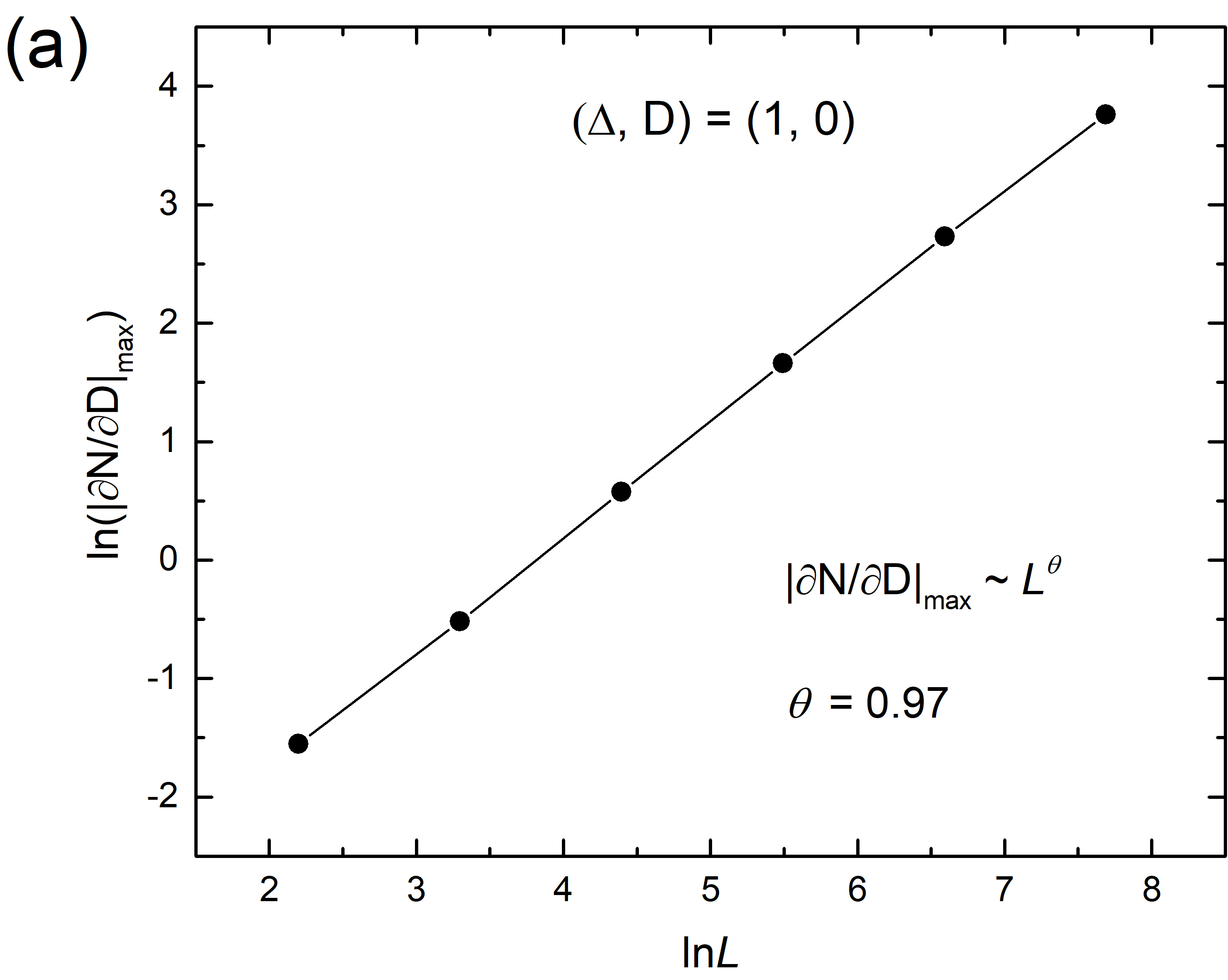}\includegraphics[width=82mm]{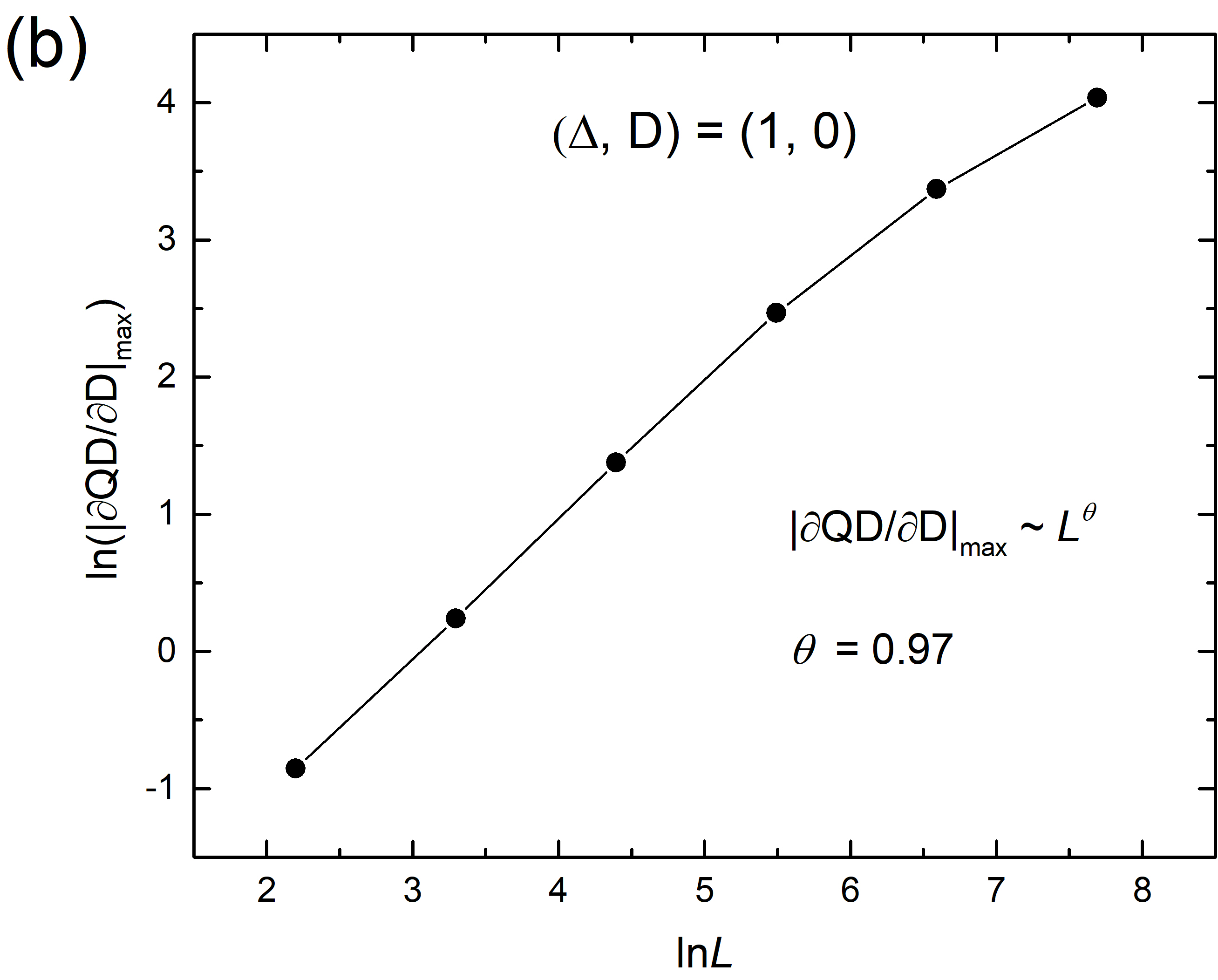}

\justifying{Figure S4. Logarithm of the maximum of (a) $\left\vert \partial N/\partial D\right\vert $
and (b) $\left\vert \partial QD/\partial D\right\vert $ versus the
logarithm of the system size, $lnL$, at $P_{1}:(1.0,0)$.}
\end{figure}

\begin{figure}[ht]
\centering\includegraphics[width=82mm]{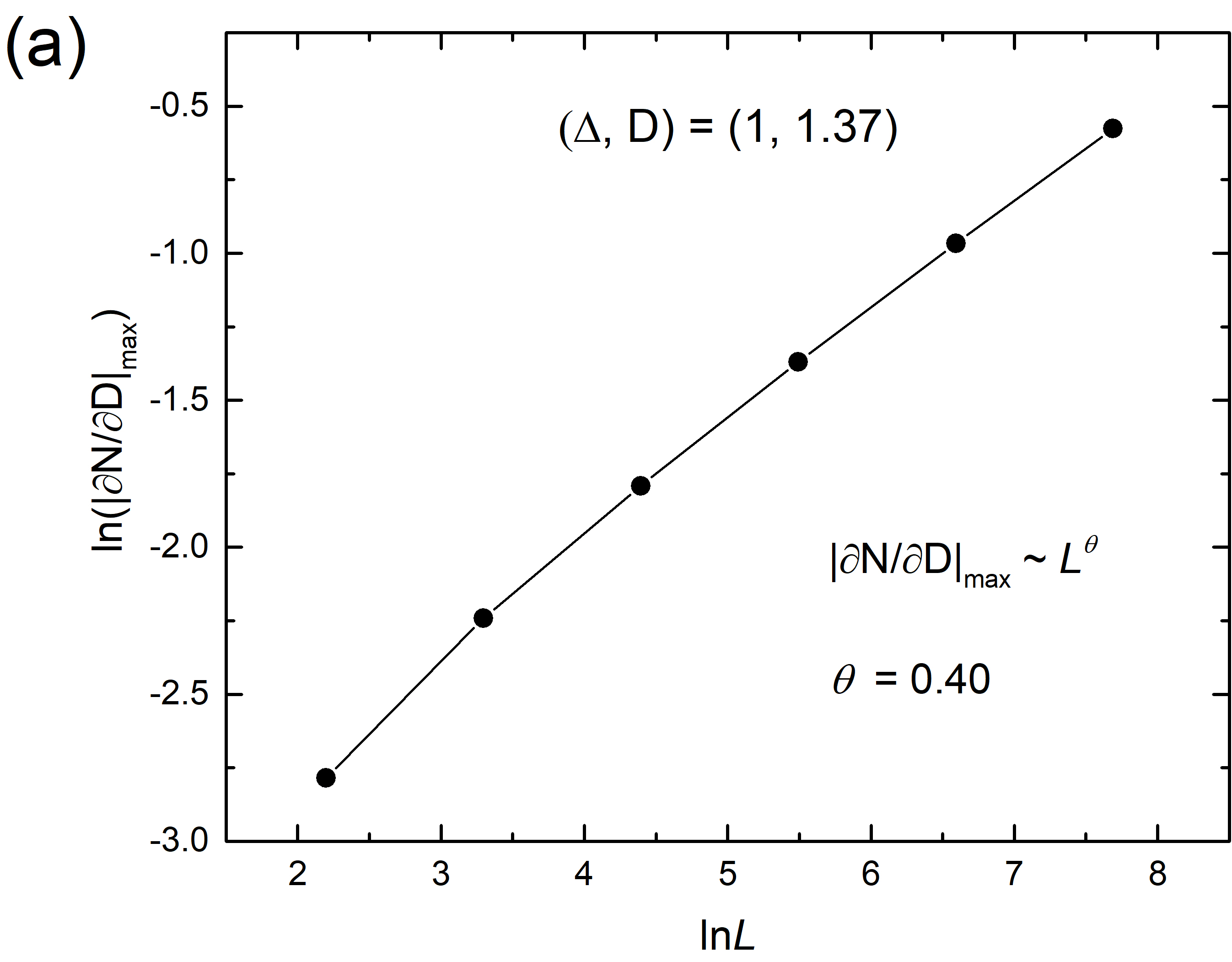}\includegraphics[width=82mm]{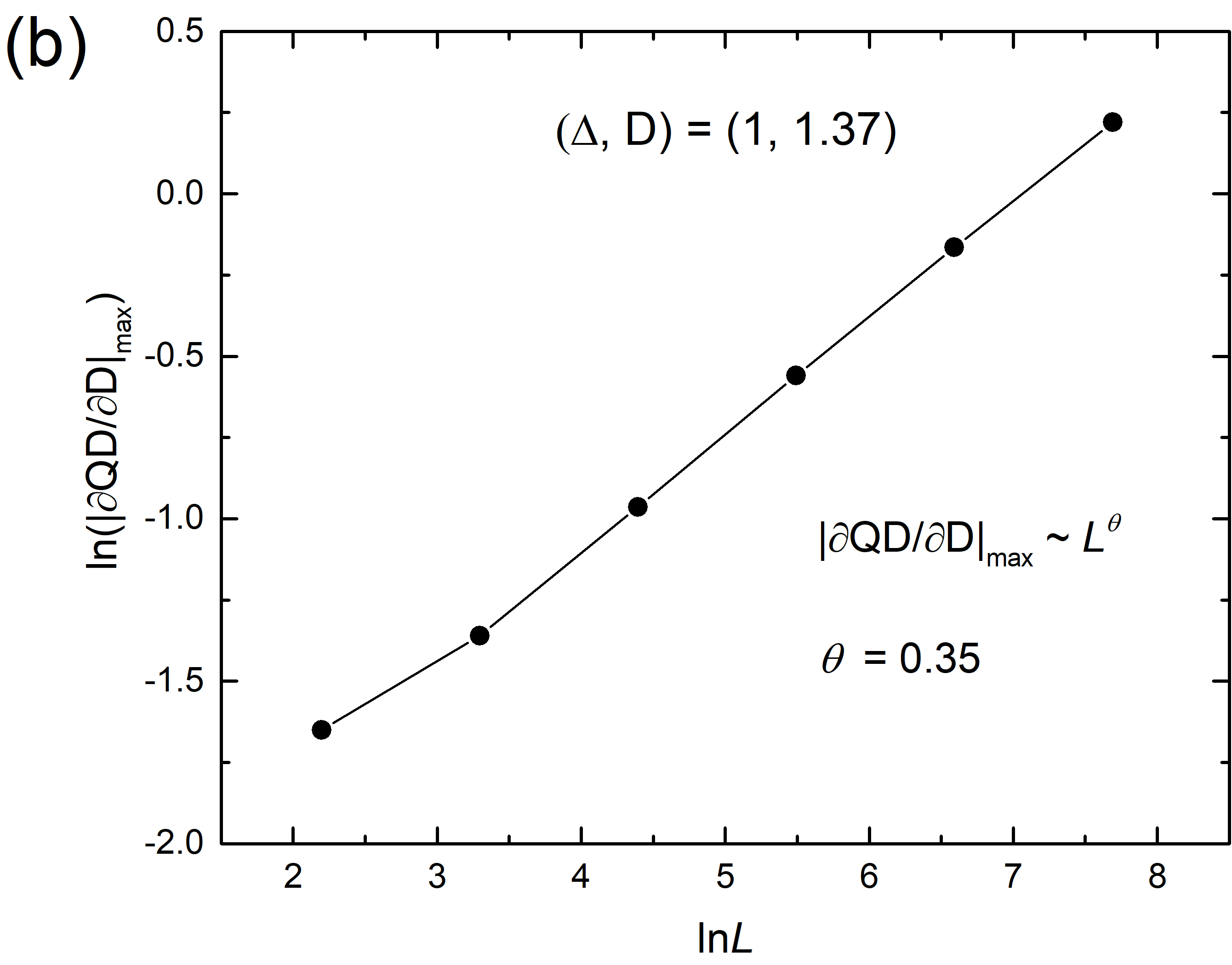}

\justifying{Figure S5. Logarithm of the maximum of (a) $\left\vert \partial N/\partial D\right\vert $
and (b) $\left\vert \partial QD/\partial D\right\vert $ versus the
logarithm of the system size, $lnL$, at $P_{6}:(1.0,1.37)$.}
\end{figure}

\begin{figure}[ht]
\centering\includegraphics[width=82mm]{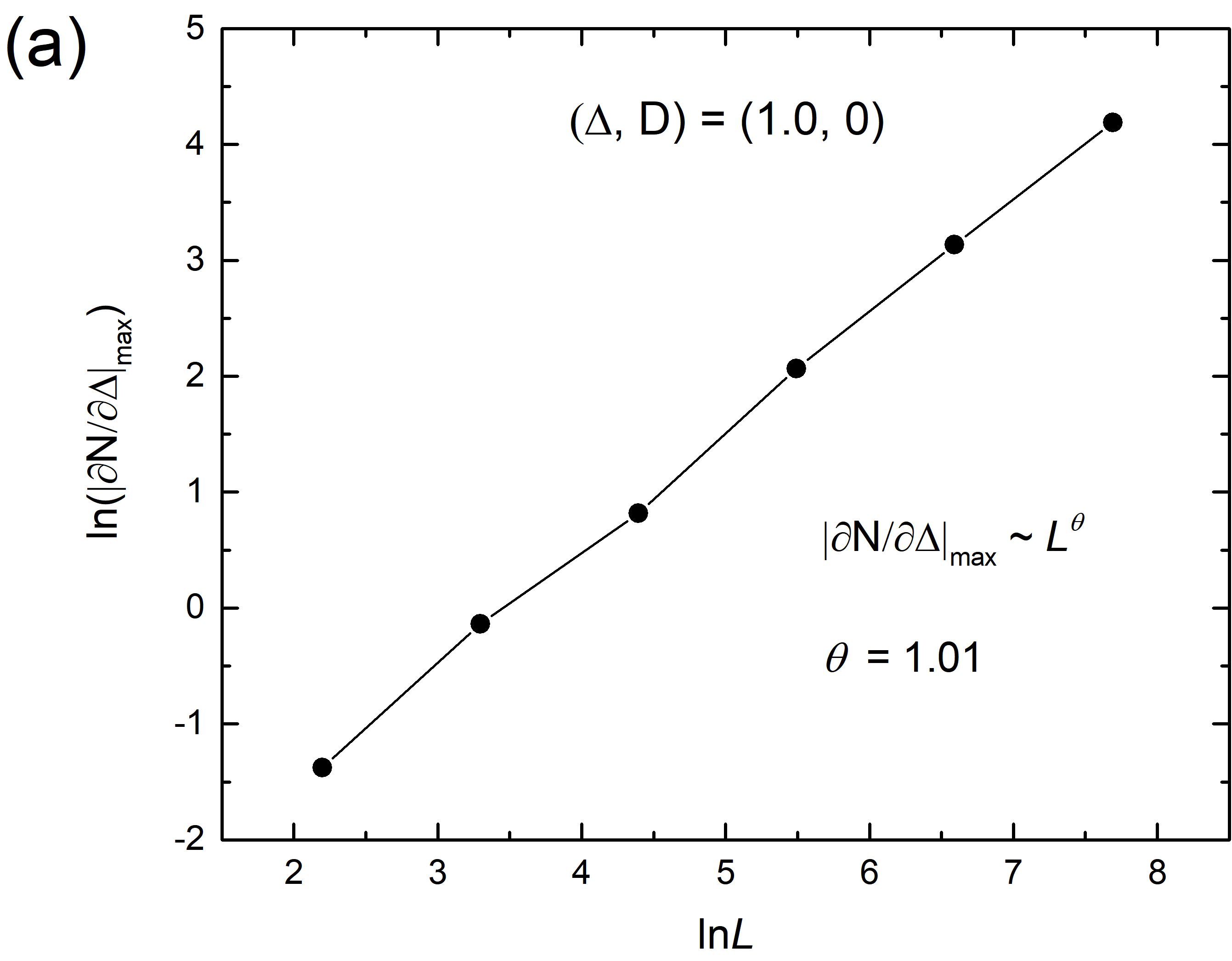}\includegraphics[width=82mm]{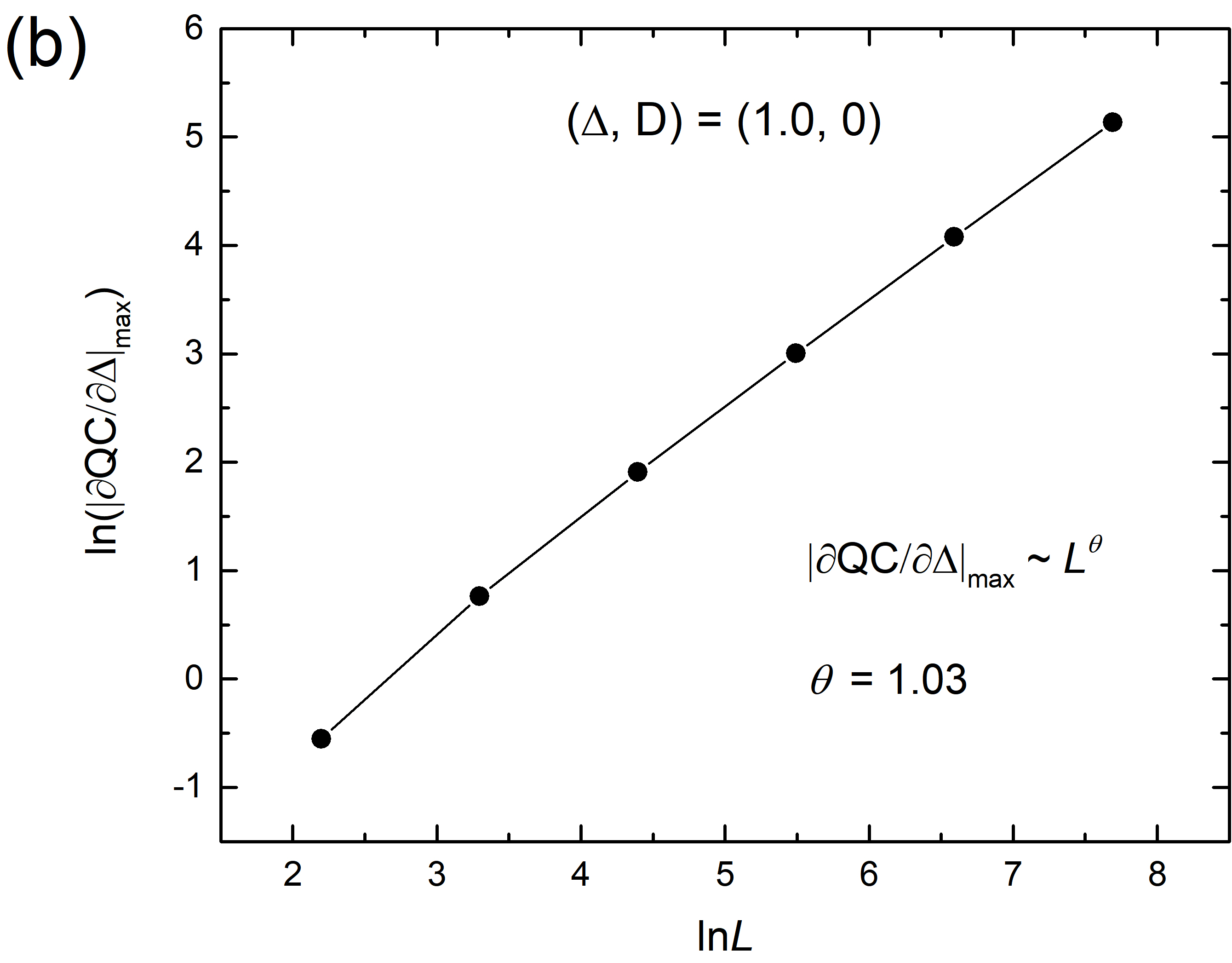}

\justifying{Figure S6. Logarithm of the maximum of (a) $\left\vert \partial N/\partial\Delta\right\vert $
and (b) $\left\vert \partial QD/\partial\Delta\right\vert $ versus
the logarithm of the system size, $lnL$, at $P_{1}:(1.0,0)$.}
\end{figure}

\begin{figure}[ht]
\centering\includegraphics[width=82mm]{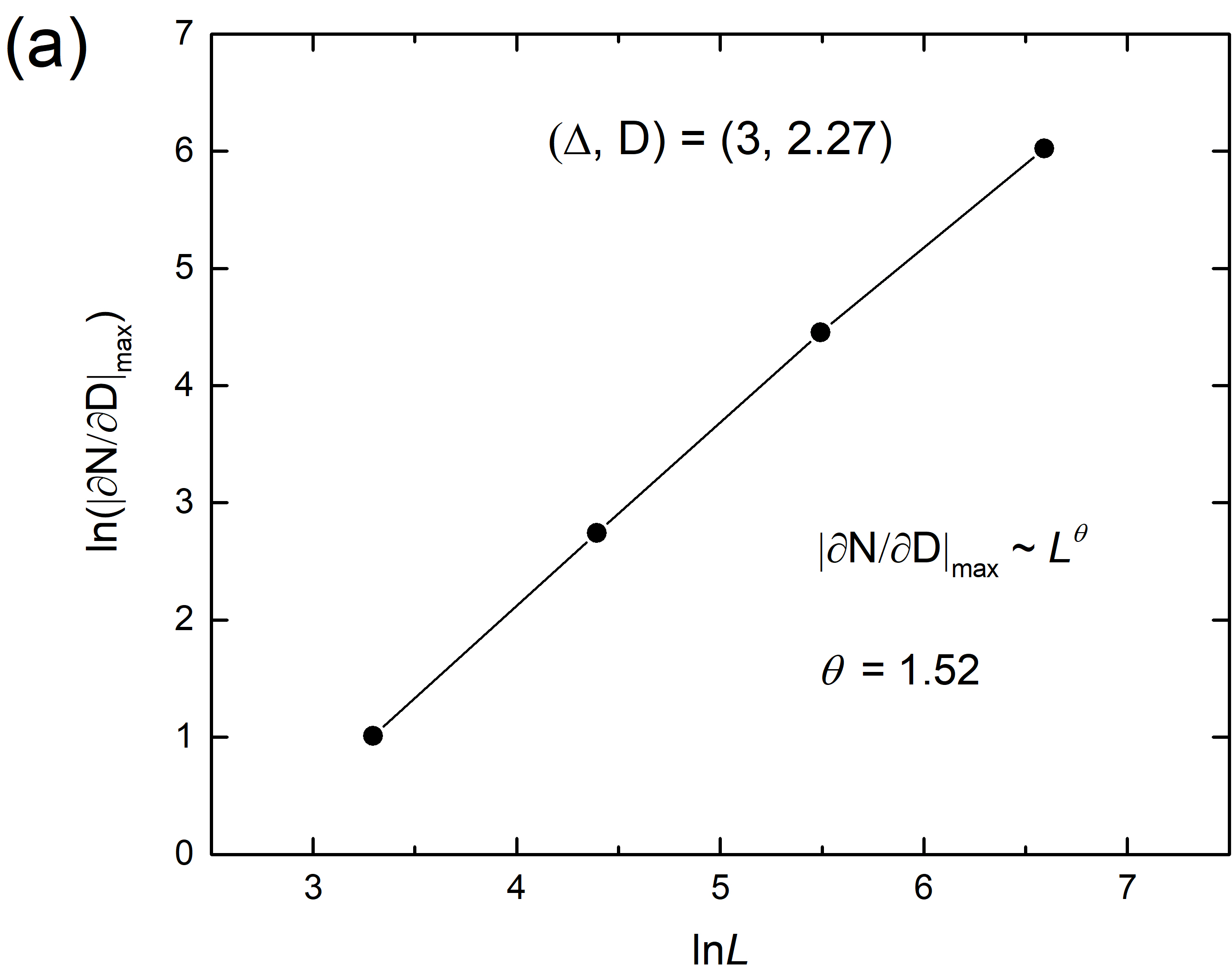}\includegraphics[width=82mm]{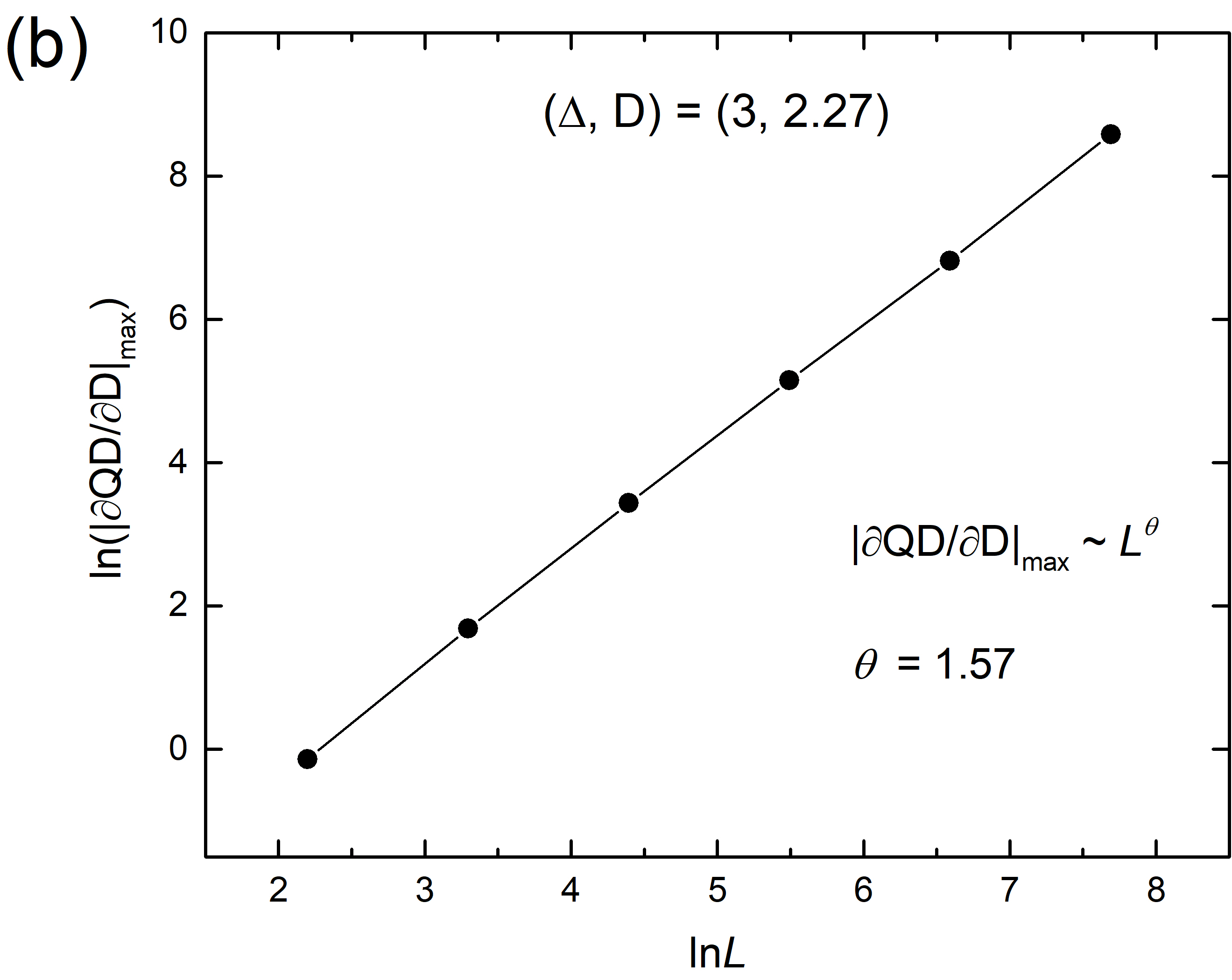}

\justifying{Figure S7. Logarithm of the maximum of (a) $\left\vert \partial N/\partial D\right\vert $
and (b) $\left\vert \partial QD/\partial D\right\vert $ versus the
logarithm of the system size, $lnL$, at $P_{5}:(3.0,2.27)$.}
\end{figure}

\clearpage{}  \bibliographystyle{elsarticle-num}
\bibliography{spin}

\end{document}